\documentclass[12pt, letterpaper]{JHEP3}

\usepackage{amssymb, amsmath, amsopn, amsthm}

\usepackage{epsfig}

\newcommand{\eref}[1]{(\ref{#1})}

\newcommand{\fref}[1]{Figure~\ref{#1}}
\newcommand{\cref}[1]{Chapter~\ref{#1}}
\newcommand{\beq}{\begin{equation}}
\newcommand{\eeq}{\end{equation}}
\newcommand{\ba}{\begin{array}}
\newcommand{\ea}{\end{array}}
\newcommand{\bcenter}{\begin{center}}
\newcommand{\ecenter}{\end{center}}

\def\IB{\relax\hbox{$\inbar\kern-.3em{\rm B}$}}
\def\IC{\relax\hbox{$\inbar\kern-.3em{\rm C}$}}
\def\ID{\relax\hbox{$\inbar\kern-.3em{\rm D}$}}
\def\IE{\relax\hbox{$\inbar\kern-.3em{\rm E}$}}
\def\IF{\relax\hbox{$\inbar\kern-.3em{\rm F}$}}
\def\IG{\relax\hbox{$\inbar\kern-.3em{\rm G}$}}
\def\IGa{\relax\hbox{${\rm I}\kern-.18em\Gamma$}}
\def\IH{\relax{\rm I\kern-.18em H}}
\def\IK{\relax{\rm I\kern-.18em K}}
\def\IL{\relax{\rm I\kern-.18em L}}
\def\IP{\relax{\rm I\kern-.18em P}}
\def\IR{\relax{\rm I\kern-.18em R}}
\def\IZ{\relax\ifmmode\mathchoice
{\hbox{\cmss Z\kern-.4em Z}}{\hbox{\cmss Z\kern-.4em Z}}
{\lower.9pt\hbox{\cmsss Z\kern-.4em Z}}
{\lower1.2pt\hbox{\cmsss Z\kern-.4em Z}}\else{\cmss Z\kern-.4em Z}\fi}
\def\II{\relax{\rm I\kern-.18em I}}

\def\sCC{{\kern 0.27em\vrule height1.45ex width0.03em depth0em
          \kern-0.30em\rm C}}
\def\C{{\mathchoice
  {\sCC}
  {\sCC}
  {\kern 0.225em \vrule height1.05ex width0.025em depth0em \kern-0.25em \rm C}
  {\kern 0.180em \vrule height0.78ex width0.02em depth0em \kern-0.2em \rm C}
        }}
\def\sHH{{\rm I\kern-.16em{}H}}
\def\H{{\mathchoice
  {\sHH}
  {\sHH}
  {\rm I\kern-.13em{}H}
  {\rm I\kern-.13em{}H} }}
\def\sNN{{\rm I\kern-.16em{}N}}
\def\N{{\mathchoice
  {\sNN}
  {\sNN}
  {\rm I\kern-.12em{}N}
  {\rm I\kern-.10em{}N} }}
\def\sPP{{\rm I\kern-.16em{}P}}
\def\P{{\mathchoice
  {\sPP}
  {\sPP}
  {\rm I\kern-.12em{}P}
  {\rm I\kern-.10em{}P} }}
\def\sQQ{{\kern 0.27em \vrule height1.45ex width0.03em depth0em
          \kern-0.30em \rm Q}}
\def\Q{{\mathchoice
        {\sQQ}
        {\sQQ}
  {\kern 0.225em \vrule height1.05ex width0.025em depth0em \kern-0.25em \rm Q}
  {\kern 0.180em \vrule height0.78ex width0.020em depth0em \kern-0.20em \rm Q}
        }}
\def\sRR{{\rm I\kern-0.16em{}R}}
\def\R{{\mathchoice
  {\sRR}
  {\sRR}
  {\rm I\kern-0.12em{}R}
  {\rm I\kern-0.10em{}R} }}
\def\sZZ{{\rm Z\kern-0.32em{}Z}}
\def\Z{{\mathchoice
  {\sZZ}
  {\sZZ} 
  {\rm Z\kern-0.3em{}Z}     %.3
  {\rm Z\kern-0.25em{}Z} }}  %.25
\def\ZZZ{{\rm Z\kern-0.24em{}Z}}
\def\sII{{\rm I\kern-0.16em{}I}}
\def\I{{\mathchoice
  {\sII}
  {\sII}
  {\rm I\kern-0.12em{}I}
  {\rm I\kern-0.10em{}I} }}

\def\inbar{\,\vrule height1.5ex width.4pt depth0pt}
\font\cmss=cmss10 \font\cmsss=cmss10 at 7pt

\def\smiley{\hbox{\large$\bigcirc$\hspace{-0.80em}\raise.2ex
\hbox{$\cdot\cdot$}\kern-.61em\lower.2ex\hbox{\scriptsize$\smile$}}\ }
\def\frowny{\hbox{\large$\bigcirc$\hspace{-0.80em}\raise.2ex
\hbox{$\cdot\cdot$}\kern-.635em\lower.2ex\hbox{\scriptsize$\frown$}}\ }

\def\I{{\rlap{1} \hskip 1.6pt \hbox{1}}}

\makeatletter
\let\hangafter\@hangfrom
\makeatother

\newcommand{\drawsquare}[2]{\hbox{%
\rule{#2pt}{#1pt}\hskip-#2pt%  left vertical
\rule{#1pt}{#2pt}\hskip-#1pt%  lower horizontal
\rule[#1pt]{#1pt}{#2pt}}\rule[#1pt]{#2pt}{#2pt}\hskip-#2pt%  upper horizontal
\rule{#2pt}{#1pt}}% right vertical

\newcommand{\fund}{\raisebox{-.5pt}{\drawsquare{6.5}{0.4}}}%  fund
\def\makeatletter{\catcode`\@=11}% 11:letter
\makeatletter
\def\mathbox#1{\hbox{$\m@th#1$}}%
\def\math@ccstyles#1#2#3#4#5#6#7{{\leavevmode
     \setbox0\mathbox{#6#7}%
     \setbox2\mathbox{#4#5}%
     \dimen@ #3%
     \baselineskip\z@\lineskiplimit#1\lineskip\z@
     \vbox{\ialign{##\crcr
            \hfil \kern #2\box2 \hfil\crcr
            \noalign{\kern\dimen@}%
            \hfil\box0\hfil\crcr}}}}
\def\mathaccstyles{\math@ccstyles\maxdimen}
\def\maththroughstyles{\math@ccstyles{-\maxdimen}}
\def\unity%
{\maththroughstyles{.45\ht0}\z@\displaystyle {\mathchar"006C}\displaystyle 1}
\title{M2-branes on Orbifolds of the Cone over $Q^{1,1,1}$}
\author{Sebasti\'an Franco$^1$, Igor R. Klebanov$^{2,3}$ and Diego Rodr\'iguez-G\'omez$^{2,4}$

\\

\vspace{0.2cm}
~\\

$^1$KITP, University of California, Santa Barbara, CA 93106-4030, USA \\
\vspace{0.2cm}

$^2$Joseph Henry Laboratories, Princeton University, Princeton, NJ 08544, USA \\
\vspace{0.2cm}

$^3$Princeton Center for Theoretical Science, Princeton University \\
Princeton, NJ 08544, USA \\
\vspace{0.2cm}

$^4$Center for Research in String Theory, Queen Mary University of
    London \\
 Mile End Road, London, E1 4NS, UK 
\vspace{0.2cm}

\email{sfranco@kitp.ucsb.edu, klebanov@Princeton.EDU, drodrigu@Princeton.EDU}\\

}

\abstract{We study the ${\cal N}=2$ supersymmetric Chern-Simons quiver gauge theory recently introduced in arXiv:0809.3237 to describe M2-branes on a cone over the well-known Sasaki-Einstein manifold $Q^{1,1,1}$. For Chern-Simons levels $(k, k, -k, -k)$ we argue that this theory is dual to $AdS_4\times Q^{1,1,1}/\mathbb{Z}_k$. We derive the $\mathbb{Z}_k$ orbifold action and show that it preserves geometrical symmetry $U(1)_R\times SU(2)\times U(1)$,
in agreement with the symmetry of the gauge theory. We analyze
the simplest gauge invariant chiral operators, and show that they match Kaluza-Klein harmonics on $AdS_4\times Q^{1,1,1}/\mathbb{Z}_k$. This provides a test of the gauge theory, and in particular of its sextic superpotential which plays an important role in restricting the spectrum of chiral operators. We proceed to study other quiver gauge theories corresponding to more complicated orbifolds of $Q^{1,1,1}$. In particular, we propose {\it two} $U(N)^4$ Chern-Simons gauge theories whose quiver diagrams are the same as in the 4d theories describing D3-branes on a complex cone over $F_0$, a $\mathbb{Z}_2$ orbifold of the conifold (in 4d the two quivers are related by the Seiberg duality). The manifest symmetry of these gauge theories is $U(1)_R\times SU(2)\times SU(2)$. We argue that these gauge theories at levels $(k,k,-k,-k)$ are dual to $AdS_4\times Q^{2,2,2}/\mathbb{Z}_k$. We exhibit calculations of the moduli space and of the chiral operator spectrum which provide support for this conjecture. We also briefly discuss a similar correspondence for $AdS_4\times M^{3,2}/\mathbb{Z}_k$. Finally, we discuss resolutions of the cones and their dual gauge theories.
}
\preprint{NSF-KITP-09-22 \\ PUPT-2294}

\def\be{\begin{equation}}

\def\ee{\end{equation}}

\def\bea{\begin{eqnarray}}

\def\eea{\end{eqnarray}}

\newcommand{\cN}{\mathcal{N}}

\begin{document}

\tableofcontents

%=====================================================================
\section{Introduction and summary}
%=====================================================================

Considerable progress in understanding coincident M2-branes is taking place, following the discovery by Bagger and Lambert \cite{Bagger:2006sk, Bagger:2007jr, Bagger:2007vi}, and by Gustavsson \cite{Gustavsson:2007vu}, of the 3-dimensional superconformal Chern-Simons theory with the maximal $\cN=8$ supersymmetry (these papers were inspired in part by the ideas of \cite{Basu:2004ed,Schwarz:2004yj}). The Bagger-Lambert-Gustavsson (BLG) 3-algebra construction was, under the assumption of manifest unitarity, limited to the gauge group $SO(4)$. This BLG theory is conveniently reformulated as an $SU(2)\times SU(2)$ gauge theory with conventional Chern-Simons terms having opposite levels $k$ and $-k$ \cite{VanRaamsdonk:2008ft,Bandres:2008vf}. For $k=2$ this model is believed to describe two M2-branes on the orbifold $\mathbb{R}^8/\mathbb{Z}_2$ \cite{Lambert:2008et,Distler:2008mk}, but for other values of $k$ its interpretation is less clear. A different approach to Chern-Simons matter theories with extended supersymmetry was introduced in \cite{Gaiotto:2008sd,Hosomichi:2008jd}. Aharony, Bergman, Jafferis and Maldacena (ABJM) \cite{Aharony:2008ug} proposed that $N$ M2-branes placed at the singularity of $\mathbb{R}^8/\mathbb{Z}_k$ are described by a $U(N)\times U(N)$ Chern-Simons gauge theory with levels $k$ and $-k$ (curiously, the matter content and superpotential of this theory are the same as for $N$ D3-branes on the conifold \cite{Klebanov:1998hh}). The $\mathbb{Z}_k$ group acts by simultaneous rotation in the four planes; for $k>2$ this orbifold preserves only $\cN=6$ supersymmetry. ABJM gave strong evidence that their Chern-Simons gauge theory indeed possesses this amount of supersymmetry, and further work in \cite{Benna:2008zy,Bandres:2008ry} provided confirmation of this claim. Furthermore, for $k=1,2$ the supersymmetry of the orbifold, and therefore of the gauge theory, is expected to be enhanced to $\cN=8$. This is not manifest in the classical action of ABJM theory. The symmetry enhancement for $k=1,2$ is expected to be a quantum effect due to the existence of certain `monopole operators' \cite{'tHooft:1977hy, Borokhov:2002cg,Borokhov:2003yu} which create quantized flux of a diagonal $U(1)$ magnetic field (for their recent discussions in this context, see for example \cite{Berenstein:2008dc,Klebanov:2008vq,Imamura:2009ur}).

In addition to the highly supersymmetric theories reviewed above, it is of obvious interest to
formulate AdS$_4$/CFT$_3$ dualities with smaller amounts of supersymmetry. $\cN=2$ is the smallest amount that allows for simple tests of the correspondence, due to the existence of the $U(1)_R$ symmetry, and the fact that the dimensions of short supermultiplets of operators are determined by their R-charges. The classical actions for $\cN=2$ Chern-Simons matter models are conveniently formulated using $\cN=2$ superspace (see, for example, \cite{Aharony:2008ug,Benna:2008zy, Gaiotto:2007qi}) which resembles the familiar $\cN=1$ superspace in $d=4$. Several examples of $\cN=2$ supersymmetric $AdS_4$ supergravity backgrounds have been known since the 80's (see \cite{Duff:1986hr} for a classic review).
%, and the recent progress offers new opportunities for formulating the gauge theories dual to them.
One of them is the $U(1)_R\times SU(3)$ invariant extremum \cite{Warner:1983vz} of the potential in the gauged $\cN=8$ supergravity \cite{de Wit:1982ig}, which was uplifted to an 11-dimensional warped $AdS_4$ background containing a `squashed and stretched' 7-sphere \cite{Corrado:2001nv}. In \cite{Benna:2008zy,Klebanov:2008vq} (see also \cite{Ahn:2008ya}) it was suggested that the dual gauge theory is the $k=1$ ABJM theory deformed by a superpotential term quadratic in one of the four chiral bifundamental superfields. Integrating this field out, one obtains a sextic superpotential for the remaining superfields. The Kaluza-Klein spectrum of this gauge theory matches that of the supergravity \cite{Klebanov:2008vq,Klebanov:2009kp}.

A simpler class of M-theory backgrounds are product spaces $AdS_4\times X_7$ where $X_7$ is a Sasaki-Einstein manifold \cite{Duff:1986hr}. The $\cN=2$ gauge theory dual to such a background arises on a stack of M2-branes placed at the apex of the 8-dimensional cone over $X_7$ \cite{Klebanov:1998hh}. The well-known examples of $X_7$ include the coset space $M^{3,2}$ (often called $M^{1,1,1}$) possessing $U(1)_R\times SU(3)\times SU(2)$ symmetry \cite{Witten:1981me}, and $Q^{1,1,1}$ possessing $U(1)_R\times SU(2)^3$ symmetry \cite{D'Auria:1983vy}. The Sasaki-Einstein spaces $M^{3,2}$ and $Q^{1,1,1}$ are $U(1)$ fibrations over $S^2\times CP^2$ and $S^2\times S^2\times S^2$, respectively \cite{Nilsson:1984bj,Sorokin:1984ca,Sorokin:1985ap}.
Proposals for their dual gauge theories were made 10 years ago in \cite{Fabbri:1999hw}; although they were not entirely satisfactory, they contained useful ideas and inspired further research. More recently, a very interesting set of `M-crystal' proposals was advanced in \cite{Lee:2006hw, Lee:2007kv, Kim:2007ic}, but they did not involve Chern-Simons gauge theories. During the last year, related proposals have been made in the context of $\cN=2$ Chern-Simons gauge theory. A proposal \cite{Martelli:2008si,Hanany:2008cd} for the theory dual to $AdS_4\times M^{3,2}$ involves a $U(N)^3$ gauge theory with levels $(-2,1,1)$; the matter content and cubic superpotential of this theory are the same as for $N$ D3-branes on $C^3/\mathbb{Z}_3$. The global symmetry of the gauge theory, $U(1)_R\times SU(3)\times U(1)$, is smaller than the geometrical symmetry of $M^{3,2}$. Yet, this does not necessarily invalidate the proposal: similarly to the ABJM theory with $k=1$, the global symmetry may be enhanced. A partial check on this proposal is that, for levels $(-2k,k,k)$ the moduli space corresponds to an orbifold $M^{3,2}/\mathbb{Z}_k$ whose action breaks the $SU(2)$ part of the global symmetry.

The goal of this paper is further exploration of
the proposal for a quiver Chern-Simons gauge theory dual to $AdS_4\times Q^{1,1,1}$ \cite{Franco:2008um}. This is a
$U(N)^4$ gauge theory with CS levels $(1,1,-1,-1)$ coupled to certain bi-fundamental chiral superfields endowed with a sextic super-potential; its details will be reviewed in \ref{section_general_Q111}. The moduli space of the abelian theory was calculated in \cite{Franco:2008um} and found to agree with the Calabi-Yau cone over $Q^{1,1,1}$. However, the manifest global symmetries of the gauge theory are only $U(1)_R\times SU(2)\times U(1)$, which are smaller than the geometrical symmetries of $Q^{1,1,1}$. In search of an explanation for this fact, we suggest that the gauge theory at level $k$ is dual to $AdS_4\times Q^{1,1,1}/\mathbb{Z}_k$
where the action of $\mathbb{Z}_k$ breaks the geometrical symmetry to $U(1)_R\times SU(2)\times U(1)$. Therefore, in the large $k$ limit where the gauge theory becomes weakly coupled, there is no conflict with the AdS/CFT correspondence \cite{Maldacena:1997re,Gubser:1998bc,Witten:1998qj}.\footnote{For $k=1$ we anticipate a quantum restoration of the $SU(2)^3$ global symmetry with the help of monopole operators; unfortunately, it is difficult to exhibit it explicitly.} In section 3 we study the gauge theory \cite{Franco:2008um} at level $k$, and explicitly derive the action of the $\mathbb{Z}_k$ orbifold. In section 4 the simplest chiral operators in this gauge theory are analyzed, and shown to match Kaluza-Klein harmonics on $AdS_4\times Q^{1,1,1}/\mathbb{Z}_k$. This provides a test of the gauge theory, and in particular of its sextic superpotential which plays an important role in restricting the spectrum of chiral operators.

Thus, exploration of the proposal \cite{Franco:2008um} naturally leads to orbifolds of $Q^{1,1,1}$ which preserve $\cN=2$ supersymmetry. In addition to changing the level $k$, we will consider changing the structure of the quiver gauge theory. A well-known projection technique \cite{Douglas:1996sw} has been used to generate new $AdS_5\times CFT_4$ dual pairs \cite{Kachru:1998ys,Lawrence:1998ja}. More recently, such $\mathbb{Z}_n$ projections have been applied to the BLG and ABJM theories \cite{Fuji:2008yj,Hosomichi:2008jd,Benna:2008zy}; somewhat surprisingly they lead to $\mathbb{Z}_n\times \mathbb{Z}_{kn}$ orbifolds of $AdS_4\times S^7$ as demonstrated through direct calculation of the moduli space \cite{Imamura:2008nn,Terashima:2008ba}. In section 5 we apply a $\mathbb{Z}_2$ projection to the quiver gauge theory of \cite{Franco:2008um}. We find a $U(N)^{8}$ quiver gauge theory which we conjecture to be dual to the $AdS_4\times Q^{1,1,1}/(\mathbb{Z}_2\times \mathbb{Z}_{2k})$ background. This conjecture is given partial support through moduli space calculations, which we present in Appendix A.

Yet another $\cN=2$ preserving orbifold of $Q^{1,1,1}$ is the space $Q^{2,2,2}=Q^{1,1,1}/\mathbb{Z}_2$ obtained through reducing the length of the $U(1)$ fiber by a factor of $2$ (reducing it by a bigger factor produces spaces $Q^{p,p,p}$, $p>2$, which turn out to break all supersymmetry).\footnote{We thank M. Benna for discussions on this issue.} We find that this kind of projection on the gravity side does not obviously correspond to a projection of the theory \cite{Franco:2008um}. Instead, in section 6 we propose two different $U(N)^4$ quiver gauge theories as candidate duals for $AdS_4\times Q^{2,2,2}$. Our proposals rely on the connections between 3d and 4d quiver gauge theories which were first observed in \cite{Aharony:2008ug} (the gauge group, matter content and superpotential of the ABJM theory are the same as in the 4d gauge theory for D3-branes on the conifold \cite{Klebanov:1998hh}), and later extended and generalized in \cite{Martelli:2008si,Imamura:2008nn,Hanany:2008cd}.

Analogously, we propose that the gauge group, matter content and superpotential of the theory dual to $AdS_4\times Q^{2,2,2}$ are the same as for the $\mathbb{Z}_2$ orbifold of the conifold theory called the $F_0$ theory \cite{Morrison:1998cs}. We study two versions of CS gauge theories with levels $(k,k,-k,-k)$; their quiver diagrams are related by the 4d Seiberg duality \cite{Seiberg:1994pq}.\footnote{One of these quiver diagrams has already made an appearance in \cite{Benna:2008zy} as a $\mathbb{Z}_2$ projection of the ABJM theory. However, in that case the choice of CS levels, $(k,-k,k,-k)$, is different from the one in the present paper.}
We conjecture that they are dual to M-theory on
$AdS_4\times Q^{2,2,2}/\mathbb{Z}_k$; in this case the $\mathbb{Z}_k$ breaks the global symmetry to $U(1)_R\times SU(2)^2$. We provide some support for this conjecture by analyzing the simplest chiral operators in the gauge theory and matching them with Kaluza-Klein harmonics.
In section 7 we make a small detour and discuss a similar operator matching for $AdS_4\times M^{3,2}/\mathbb{Z}_k$.
Finally, in section 8 we consider giving vacuum expectation values to some of the chiral superfields, and compare this with placing the M2-branes on resolved cones.

\bigskip
\noindent {\bf Note added:} after this paper was written, the authors of \cite{Hanany} and \cite{Forcella} informed us of their upcoming work, in which 3d CS quivers are also studied.

%=====================================================================
\section{$Q^{1,1,1}$ and its dual gauge theory}
%=====================================================================

\label{section_general_Q111}

$Q^{1,1,1}$ is the homogenous coset space
\beq
{SU(2) \times SU(2) \times SU(2)\over U(1) \times U(1)} \,
\eeq
which has $U(1)_R\times SU(2)^3 $ isometry \cite{D'Auria:1983vy}. Its metric is conveniently written as a $U(1)$ bundle over
$S^2\times S^2\times S^2$ \cite{Nilsson:1984bj,Sorokin:1984ca,Sorokin:1985ap}
\begin{equation}
ds^2_{Q^{1,1,1}}=\frac{1}{16}\big(d\psi+\sum_{i=1}^3\cos\theta_id\phi_i\big)^2+\frac{1}{8}\sum_{i=1}^3\big(d\theta_i^2+\sin^2\theta_id\phi_i^2\big)\ ,
\label{metric_Q111}
\end{equation}
with $\theta_i \in [0,\pi)$, $\phi_i \in [0,2\pi)$ and $\psi \in [0,4\pi)$.
The cone\footnote{Throughout the paper, we use the notation ${\mathcal C}(X_7)$ to denote the 8 real dimensional cone with 7-dimensional base $X_7$.} over $Q^{1,1,1}$  has metric $dr^2 + r^2 ds^2_{Q^{1,1,1}}$; it is a Calabi-Yau 4-fold with holomorphic 4-form
{\footnotesize
\begin{equation}
\Omega \sim r^4 e^{i\psi}\Big(\frac{dr}{r}+\frac{i}{4}\big(d\psi+\sum\cos\theta_id\phi_i\big)\Big)\wedge\Big(d\theta_1+i\sin\theta_1d\phi_1\Big) \Big(d\theta_2+i\sin\theta_2d\phi_2\Big)\wedge\Big(d\theta_3+i\sin\theta_3d\phi_3\Big)\, .
\label{holomorphic_4-form}
\end{equation}}

\begin{figure}[h]
\begin{center}
\includegraphics[width=10cm]{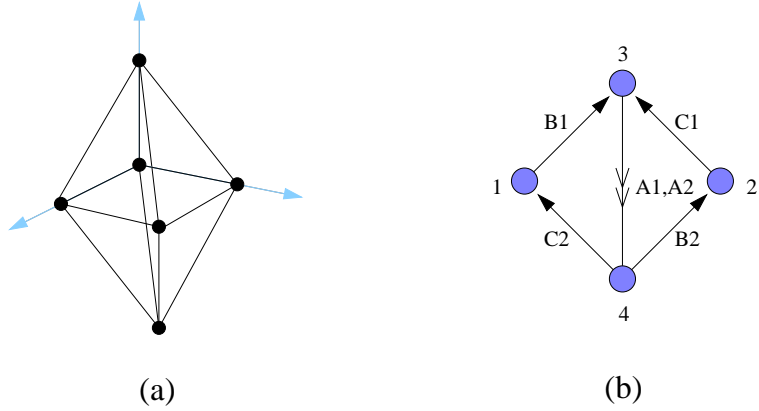}
\caption{(a) toric diagram and (b) proposed quiver diagram for ${\cal C}(Q^{1,1,1})$.}
\label{toric_quiver_Q111}
\end{center}
\end{figure}

The toric diagram for ${\mathcal C}(Q^{1,1,1})$ is shown in \fref{toric_quiver_Q111}.a. Its toric geometry is described in terms of three $SU(2)$ doublets of complex coordinates: $(A_1, A_2)$, $(B_1, B_2)$, $(C_1, C_2)$. The $SU(2)^3$ symmetry is manifest in this description, but these coordinates are not gauge invariant. The 8 gauge invariant combinations are \cite{Fabbri:1999hw,Herzog:2000rz}
\beq
\begin{array}{lclclcl}
w_1 = A_1B_2C_1 & \ \ \ & w_2 = A_2B_1C_2 & \ \ \ &  w_3 = A_1B_1C_2 & \ \ \ & w_4 = A_2B_2C_1 \\
w_5 = A_1B_1C_1 & \ \ \ & w_6 = A_2B_1C_1 & \ \ \ & w_7 = A_1B_2C_2 & \ \ \ & w_8 = A_2B_2C_2  \, ,
\end{array}
\label{ws}
\eeq
which satisfy 9 relations

\beq
\begin{array}{cccccccc}
w_1w_2 - w_3w_4 & = & w_1w_2 - w_5w_8 & = & w_1w_2 - w_6w_7 & = & 0 & \\
w_1w_3 - w_5w_7 & = & w_1w_6 - w_4w_5 & = & w_1w_8 - w_4w_7 & = & 0 & \\
w_2w_4 - w_6w_8 & = & w_2w_5 - w_3w_6 & = & w_2w_7 - w_3w_8 & = & 0 & ,
\label{eq_Q11}
\end{array}
\eeq
%\beq
%\begin{array}{ccc}
%w_1w_2 - w_3w_4 = 0 & \ \ \ \ \ \ \ & w_1w_2 - w_6w_7 = 0 \\
%w_1w_2 - w_5w_8 = 0 & \ \ \ \ \ \ \ & w_2w_5 - w_3w_6 = 0
%\label{eq_Q11}
%\end{array}
%\eeq
describing the embedding of ${\cal C}(Q^{1,1,1})$ in $\mathbb{C}^8$.

A quiver $U(N)^4$ CS gauge theory for M2-branes probing ${\cal C}(Q^{1,1,1})$  was proposed in \cite{Franco:2008um}.
As usual, the coordinates $A_i, B_j, C_l$ were promoted to bifundamental chiral superfields (this was also proposed in \cite{Fabbri:1999hw} but there the gauge group was only $U(N)^3$).
The quiver diagram of \cite{Franco:2008um} is shown in \fref{toric_quiver_Q111}.b and its superpotential is
\begin{equation}
W = {\rm Tr} (C_2 \, B_1 \, A_1 \, B_2 \, C_1 \, A_2 - C_2 \, B_1 \, A_2 \, B_2 \, C_1 \, A_1) \,.
\label{W_Q111}
\end{equation}
%

%======================================================================
%======================================================================

The quiver and superpotential have a manifest $SU(2)_1$ global symmetry under
which the chiral fields $A_i$ form a doublet. The marginality of the superpotential imposes constraints on the R-charges:
\beq \label{constraint}
R(A_i)+ R(B_2)+ R(C_1)= R(A_i)+ R(B_1) + R(C_2) =1
\ .
\eeq
In addition, there is a non-R $U(1)$ symmetry; we assign the following charges
under this symmetry
\beq \label{globalcharge}
Q(B_1)={1\over 2}\ , \quad
Q(B_2)=-{1\over 2}\ , \quad   Q(C_1)=-{1\over 2}\ , \quad
Q(C_2)={1\over 2}\ .
\eeq

The CS levels are $\vec{k}=(k,k,-k,-k)$. In \cite{Franco:2008um}, the moduli space of the abelian $N=1$ gauge theory with $k=1$
was computed using toric geometry techniques and shown to correspond to ${\cal C}(Q^{1,1,1})$.\footnote{In \cite{Franco:2008um}, it was also shown that the choice $\vec{k}=(1,-1,0,0)$ leads
to the same moduli space.} This provided a test of the theory proposed in \cite{Franco:2008um}.

On the other hand, the $U(1)_R\times SU(2)\times U(1)$ symmetry of the non-abelian gauge theory is only a subset of the $U(1)_R\times SU(2)^3$ geometrical symmetry of
$Q^{1,1,1}$. This is an important difference from the early proposal \cite{Fabbri:1999hw} which also attempted to introduce bi-fundamental chiral superfields $A_1, A_2, B_1, B_2, C_1, C_2$ and a sextic superpotential for them. However, it seemed impossible to write down such a quiver gauge theory with manifest $U(1)_R\times SU(2)^3$. The proposal of \cite{Franco:2008um} circumvents this problem by reducing the manifest symmetry. In the next section we will argue that, for $k>1$, the gauge theory is actually dual to $AdS_4\times Q^{1,1,1}/\mathbb{Z}_k$, and that the orbifold action explains the reduction of symmetry to $U(1)_R\times SU(2)\times U(1)$.

%=====================================================================
\section{The $Q^{1,1,1}$ gauge theory at higher CS level}
%=====================================================================

\label{section_higher_CS}

In a general CS quiver gauge theory, one may define $k={\rm gcd}(k_i)$. When passing from $k=1$ to arbitrary $k$, the moduli space of a quiver CS theory changes from $\mathcal{M}$ (the one arising solely from F and D-flatness) to $\mathcal{M}/\mathbb{Z}_k$. This follows from a by now standard argument \cite{Imamura:2008nn,Martelli:2008si,Hanany:2008cd} that we now review. Denote by $n_G$ the number of gauge groups. Let us consider the abelian $U(1)^{n_G}$ theory. For each node, we denote the corresponding gauge field as $\mathcal{A}_i$. It is straightforward to see that the overall $U(1)$ given by $\mathcal{B}_{n_G}=\sum_i\mathcal{A}_i$ is decoupled from the scalars, since all of them transform in bifundamental or adjoint representations. This field only appears through the CS coupling
\begin{equation}
\label{SB4}
S(\mathcal{B}_{n_G})=\frac{k}{n_G 2\pi}\int\, (\mathcal{B}_{n_G-1})_{\mu}\epsilon^{\mu\nu\rho}(\mathcal{G}_{n_G})_{\nu\rho}\ ,
\end{equation}
where $\mathcal{G}_{n_G}=d\mathcal{B}_{n_G}$ and
\begin{equation}
\mathcal{B}_{n_G-1}=\frac{1}{k} \sum_i k_i\mathcal{A}_i \, .
\end{equation}
We can dualize $\mathcal{B}_{n_G}$ into a scalar. We interpret $\mathcal{G}_{n_G}$ as an independent variable and add a Lagrange multiplier imposing $\mathcal{G}_{n_G}=d\mathcal{B}_{n_G}$:
\begin{equation}
\label{Stau}
S(\tau)=\frac{1}{2\pi}\int \, \tau\epsilon^{\mu\nu\rho}\partial_{\mu}(\mathcal{G}_{n_G})_{\nu\rho}\, .
\end{equation}
Using the equations of motion for $\mathcal{G}_{n_G}$, we have
\begin{equation}
\label{eom}
(\mathcal{B}_{n_G-1})_{\mu}=\frac{n_G}{k}\partial_{\mu}\tau\, .
\end{equation}
Taking the full action for this sector (\ref{SB4})+(\ref{Stau}), integrating it by parts and using (\ref{eom}), we get
\begin{equation}
S=\int\,\partial_{\mu}\Big(\frac{\tau}{2\pi}\epsilon^{\mu\nu\rho}(\mathcal{G}_{n_G})_{\nu\rho}\Big)\ .
\end{equation}
This is a total derivative; however, in order for this phase to be unobservable, $\tau$ must be a periodic variable with period $2\pi/n_G$. Following \cite{Imamura:2008nn,Martelli:2008si}, we impose $\int \star\mathcal{G}_{n_G}= 2 \pi n n_G$.\footnote{This can be argued to follow from the original CS normalization. Since we are normalizing the CS action with $1/4\pi$ for each $\mathcal{A}_i$, we are implicitly assuming that $\int \star \mathcal{F}_i=2\pi$, so given the definition of $\mathcal{B}_{n_G}$ it seems reasonable to assume the normalization we chose.}

We can now go back to (\ref{eom}), and note that we can locally set $\tau$ to a constant by $\mathcal{B}_{n_G-1}$ gauge transformations. However, the large gauge transformations for $\mathcal{B}_{n_G-1}$ inherit the periodicity of $\tau$. Indeed, if we call the parameter of these transformations $\Lambda_{n_G-1}$, we have $\Lambda_{n_G-1}=\frac{2\pi}{k}$. More explicitly, $\Lambda_{n_G-1}=k^{-1}\sum_i k_i\theta_i$, where $\theta_i$ is the gauge parameter for the $i$-th node.

Let us now focus on the gauge transformations orthogonal to $\mathcal{B}_{n_G}$, i.e. those which leave this field unaffected. Since $\sum k_i=0$, they are of the form $\theta_i=k_i \theta$, for some constant $\theta$. After a straightforward computation, we get $\theta=\frac{2\pi}{\vec{k}^2}$.

Let us now specialize the above general discussion to the case of the $Q^{1,1,1}$ theory with CS levels $\vec{k}=(k,k,-k,-k)$. Following the expressions above, we have that $\theta_1=\theta_2=-\theta_3=-\theta_4=\frac{\pi}{2}$. The identifications imposed by the large gauge transformations on the scalar fields are
\begin{equation}
\label{identification}
(A_1,A_2)\sim(A_1,A_2)\ ,\qquad (B_1,B_2)\sim (e^{i\frac{\pi}{k}} B_1,e^{-i\frac{\pi}{k}} B_2)\ , \qquad(C_1,C_2)\sim(e^{i\frac{\pi}{k}}C_1,e^{-i\frac{\pi}{k}}C_2) \, .
\end{equation}
In terms of the angular coordinates in (\ref{metric_Q111}) this corresponds to
\beq
(\phi_2,\phi_3) \sim (\phi_2,\phi_3)+\left(\frac{2\pi}{k},\frac{2\pi}{k}\right) \, .
\eeq
Clearly, the $\mathbb{Z}_k$ orbifold does not affect the holomorphic 4-form (\ref{holomorphic_4-form}), and hence preserves the ${\cal N}=2$
supersymmetry.
However, it preserves only the $SU(2)_1\times U(1)$ subgroup of the global symmetry.
In terms of the coordinates $w_i$ (\ref{ws}), the orbifold action is given by
\begin{equation}
(w_1,w_2,w_3,w_4,w_5,w_6,w_7,w_8)\, \rightarrow (
w_1,w_2,w_3,w_4,e^{i\frac{2\pi}{k}}w_5,e^{i\frac{2\pi}{k}}w_6,e^{-i\frac{2\pi}
{k}}w_7,e^{-i\frac{2\pi}
{k}}w_8) \, .
\label{orbifold_action}
\end{equation}
This confirms that we are taking a $\mathbb{Z}_k$ orbifold of $Q^{1,1,1}$.

In the abelian $N=1$ theory,
the four operators $w_1, w_2, w_3, w_4$ from (\ref{ws}) are fully gauge invariant, while $w_5, w_6, w_7, w_8$ are only invariant with respect to $Q_1+Q_3$ and $Q_1+Q_4$, the two $U(1)$'s defined by the choice $\vec{k}=(k,k,-k,-k)$. The latter four are not invariant under the $\mathbb{Z}_k$ orbifold action. Together these operators correspond to the eight harmonics of R-charge 1 on $Q^{1,1,1}$ \cite{Merlatti:2000ed}, but only the first four correspond to allowed harmonics on $Q^{1,1,1}/\mathbb{Z}_k$. We will describe an extension of this matching to the non-abelian $N>1$ gauge theory in the next section. %\ref{section_matching_chiral_operators_Q111}.

%=====================================================================
\section{Matching of chiral operators}
%=====================================================================

\label{section_matching_chiral_operators_Q111}

An essential test of the AdS/CFT correspondence involves matching the Kaluza-Klein supergravity modes with gauge-invariant operators \cite{Gubser:1998bc,Witten:1998qj}. For $3$-dimensional theories with ${\cal N}=2$ superconformal symmetry there exist chiral operators whose dimension is given by the absolute value of the $U(1)_R$ charge. The simplest such spherical harmonics on $AdS_4\times Q^{1,1,1}$ were found in \cite{Fabbri:1999hw,Merlatti:2000ed}: in terms of the coordinates $A_i, B_j, C_l$, they are given by
\beq
\prod_{a=1}^r A_{i_a} B_{j_a} C_{l_a}\ .
\eeq
They carry $U(1)_R$-charge $r$ and transform with spins $(r/2, r/2, r/2)$ under the global $SU(2)^3$ symmetry; thus, there are
$(r+1)^3$ different harmonics.
The $\mathbb{Z}_k$ orbifold projects out some of them. For example, for $r=1$ only four out of the eight harmonics are invariant, as shown in (\ref{orbifold_action}). For $r=2$ and $k\geq 2$, only 9 out of 27 modes survive the orbifold projection:
\beq
A_i A_j B_1 B_2 C_1 C_2\ ,\qquad  A_i A_j B_1^2 C_2^2\ ,\qquad  A_i A_j B_2^2 C_1^2\ .
\eeq
In general, for R-charge $r< k$, there are $(r+1)^2$ modes invariant under the $\mathbb{Z}_k$ action; they have $SU(2)$ spin $r/2$, and the global $U(1)$ charge
$Q$, defined in (\ref{globalcharge}), ranging in integer steps from $-r$ to $r$.

Let us show that this matches the spectrum of gauge invariant operators in the quiver theory. For simplicity, we will first take $k\gg 1$ so that the theory is weakly coupled and we can ignore the monopole operators. Due to the structure of the quiver and the constraint
(\ref{constraint}), the gauge invariant mesonic operators carry integer $R$-charge $r$.
For $r=1$, there are four such operators
\beq
{\rm Tr} A_i C_2 B_1\ , \qquad  {\rm Tr} A_i B_2 C_1 \ ,
\label{ops_R1_Q111}
\eeq
and their $SU(2)\times U(1)$ charges agree with supergravity.

For $r=2$ there are 9 gauge invariant chiral operators
\beq
{\rm Tr} A_i C_2 B_1 A_j C_2 B_1\ , \qquad  {\rm Tr} A_i B_2 C_1 A_j B_2 C_1\ , \quad {\rm Tr} A_i C_2 B_1 A_j B_2 C_1\ .
\eeq
Each of these operators is symmetric under the interchange of $i$ and $j$, and thus carries $SU(2)$ spin 1.
For the first two types, this is obvious from the cyclic symmetry of the trace. For the third one it arises in a more interesting way, due to the F-term conditions coming from the superpotential:
\beq
B_1 A_1 B_2 C_1 A_2 = B_1 A_2 B_2 C_1 A_1 \ , \qquad A_2 C_2 B_1 A_1 B_2 = A_1 C_2 B_1 A_2 B_2
\eeq
Since these equations are supposed to hold for arbitrary $B_1, B_2$, they imply
\beq \label{Fterm}
A_1 B_2 C_1 A_2 = A_2 B_2 C_1 A_1 \ , \qquad A_2 C_2 B_1 A_1  = A_1 C_2 B_1 A_2 \ ,
\eeq
which means that the $A$-fields may be permuted inside operators, producing symmetry in the $SU(2)$ index. This means that each chiral operator carries only the maximum possible $SU(2)$ spin consistent with its other charges.

In general, we may define $SU(2)$ doublet operators of R-charge 1,
\beq
X_i^+ = A_i C_2 B_1\ , \qquad X_i^- = A_i B_2 C_1\ ,
\eeq
where $\pm$ denotes the $U(1)$ charge.
The R-charge $r$ chiral operators are
\beq
{\rm Tr} \prod_{a=1}^r  X^{\pm}_{i_a}
\ .\eeq
The superpotential F-term conditions (\ref{Fterm}) guarantee that the $SU(2)$ spin of such operators is $r/2$, and the $U(1)$ charges range in integer steps from $-r$ to $r$.
Therefore, as for example in the conifold gauge theory \cite{Klebanov:1998hh}, the superpotential is crucial for giving the spectrum of chiral operators matching the supergravity modes.

Let us note that for $r\geq k$ some additional supergravity modes appear that are not projected out by the orbifold. In order to construct the corresponding gauge invariant operators one would need the monopole operators, which transform non-trivially under the gauge group. Their discussion is beyond the scope of this paper.

%=====================================================================
\section{Orbifold projection of the quiver}
%=====================================================================

\label{section_Q111_Z2_Z2}

In this section we explore another simple way of orbifolding the CS $Q^{1,1,1}$ quiver theory, namely using the orbifold projection techniques of \cite{Douglas:1996sw}. There are various discrete symmetries of the gauge theory we could choose; for example, the $\mathbb{Z}_p$ symmetry
$B_1\rightarrow e^{2\pi i/p} B_1$, $B_2\rightarrow e^{-2\pi i/p} B_2$. To simplify our discussion, we will exhibit the details for the case $p=2$.

\begin{figure}[h]
\begin{center}
\includegraphics[width=5cm]{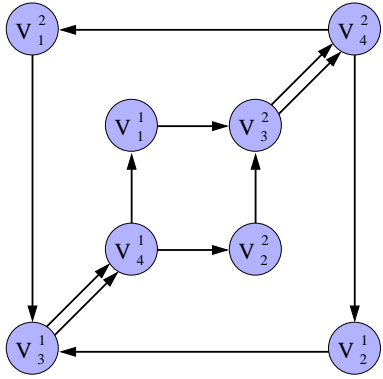}
\caption{Quiver diagram for a $\mathbb{Z}_2$ orbifold projection of the $Q^{1,1,1}$ quiver.}
\label{quiver_Q111_Z2_Z2}
\end{center}
\end{figure}

We start with the $U(2N)^4$ quiver theory and consider the $\mathbb{Z}_2$ orbifold identifications
\beq
\begin{array}{lclcl}
A_1=\Omega^{\dagger}A_1\Omega & \ \ \ \ \ & B_1=- \, \Omega^{\dagger}B_1\Omega & \ \ \ \ \ & C_1= \Omega^{\dagger}C_1\Omega \\
A_2= \Omega^{\dagger}A_2\Omega & \ \ \ \ \ & B_2=- \, \Omega^{\dagger}B_2\Omega & \ \ \ \ \ & C_2=\Omega^{\dagger}C_1\Omega \, ,
\end{array}
\eeq
where
\begin{displaymath}
\Omega=\left(\begin{array}{c c} \unity & 0\\ 0 & - \unity
\end{array}\right)
\end{displaymath}
breaks the gauge symmetry to $U(N)^8$.
We obtain
\begin{center}
\begin{tabular}{c c c}
$A_1=\left(\begin{array}{cc}
A^1_1 & 0\\
0 & A^2_1
\end{array}\right)
$
&
$B_1=\left(\begin{array}{cc}
0 & B^1_1\\
B^2_1 & 0
\end{array}\right)$
&
$C_1=\left(\begin{array}{cc}
C^1_1 & 0\\
0 & C^2_1
\end{array}\right)$
\\ & & \\
$A_2=\left(\begin{array}{cc}
A^1_2 & 0\\
0 & A_2^2
\end{array}\right)$
&
$B_2=\left(\begin{array}{cc}
0& B^1_2\\
B^2_2 & 0
\end{array}\right)$
&
$C_2=\left(\begin{array}{cc}
C^1_2 & 0\\
0 & C^2_2
\end{array}\right)$
\end{tabular}
\end{center}
The gauge fields are now

\beq
V_i=\left(\begin{array}{cc} V_i^1 & 0\\0 & V^2_i\end{array}\right)\ , i=1,2,3,4\ ,
\eeq
so the kinetic term for the chiral supermultiplets is
\begin{eqnarray}
S_{Kahler}=\int d^4\theta\, && {\rm Tr} \bigg (\bar{A}^1_i e^{-V^1_3}A^1_i e^{V^1_4}+\bar{A}^2_ie^{-V^2_3}A^2_i e^{V^2_4} + \bar{B}^2_1e^{-V^2_1}B^2_1e^{V^1_3}+\bar{B}^1_1e^{-V^1_1}B^1_1e^{V^2_3}\nonumber \\  + && \bar{B}^2_2e^{-V^2_4}B^2_2e^{V^1_2}+\bar{B}^1_2e^{-V^1_4}B^1_2e^{V^2_2}+
\bar{C}^1_1 e^{-V^1_2}C^1_1e^{V^1_3} + \bar{C}^2_1e^{-V^2_2}C^2_1 e^{V^2_3}
\nonumber\\ + && \bar{C}^1_2 e^{-V ^1_4}C^1_2e^{V^1_1}+\bar{C}^2_2e^{-V^2_4}C^2_2e^{V^2_1}\bigg ) \, .
\end{eqnarray}
This charge assignment corresponds to the quiver diagram in \fref{quiver_Q111_Z2_Z2}. The superpotential reads
\begin{equation}
W={\rm Tr} (C^1_2 \, B^1_1 \, A^2_1 \, B^2_2 \, C^1_1 \, A^1_2+C^2_2 \, B^2_1 \, A^1_1 \, B^1_2 \, C^2_1 \, A^2_2-C^1_2 \, B^1_1 \, A^2_2 \, B^2_2 \, C^1_1 \, A^1_1-C^2_2 \, B^2_1 \, A^1_2 \, B^1_2 \, C^2_1 \, A^2_1 )\, .
\end{equation}

%======================================================================
%======================================================================

We consider the choice of CS levels that descends from the parent $Q^{1,1,1}$ theory, namely $k=(1,1,1,1,-1,-1,-1,-1)$, where the order of nodes is $(V^1_1,V^2_1,V^1_2,V^2_2,V^1_3,$ $V^2_3,V^1_4,V^2_4)$.
In appendix \ref{appendix_Z2_Z2}, we compute the moduli space of this theory following \cite{Franco:2008um}. Interestingly, it is $\mathcal{C}(Q^{1,1,1}/(\mathbb{Z}_2\times \mathbb{Z}_2))$.
The ``doubling" of the orbifold group from the $\mathbb{Z}_2$ acting on the quiver to $\mathbb{Z}_2 \times \mathbb{Z}_2$ is not surprising;
the same behavior was observed in \cite{Hosomichi:2008jd,Imamura:2008nn,Terashima:2008ba} for the orbifold projections of ABJM theory introduced in \cite{Benna:2008zy}.

%=====================================================================
\section{M2-branes on ${\mathcal C}(Q^{2,2,2})$ and its orbifolds}
%=====================================================================

\label{section_Q222}

The $Q^{p,p,p}$ manifolds are $\mathbb{Z}_p$ orbifolds of $Q^{1,1,1}$ that preserve the $SU(2)^3$ isometry.
$Q^{p,p,p}$ is described by the same metric as $Q^{1,1,1}$,
(\ref{metric_Q111}), but with the $\psi$ fiber having period $4\pi/p$ The holomorphic 4-form given in \eref{holomorphic_4-form} is invariant only for $p=1$, $2$. Thus, $Q^{p,p,p}$ is supersymmetric only for $p=1$, $2$.

Our preceding analysis suggests that the gauge theory for $Q^{2,2,2}$ arises neither as the $Q^{1,1,1}$ gauge theory at CS level $2$ nor as a result of a Douglas-Moore projection of the quiver.
In this section we propose the gauge theory describing M2-branes on the cone over $Q^{2,2,2}$ and its orbifolds. Our construction is based on a correspondence with certain 4d gauge theories and gives the desired moduli space.

\begin{figure}[h]
\begin{center}
\includegraphics[width=5cm]{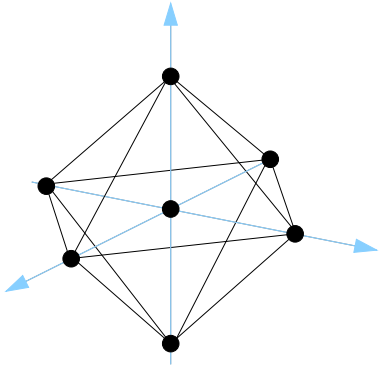}
\caption{Toric diagram for ${\cal C}(Q^{2,2,2}$).}
\label{toric_Q222_0}
\end{center}
\end{figure}

The toric diagram for ${\cal C}(Q^{2,2,2})$ is shown in \fref{toric_Q222_0}. It is a refinement of the ${\cal C}(Q^{1,1,1})$ toric diagram in \fref{toric_quiver_Q111}.a by the addition of a single internal point. This tells us that it is a $\mathbb{Z}_2$ orbifold of ${\cal C}(Q^{1,1,1})$. Furthermore, we can see it has an $SU(2)^3$ isometry by computing the GLSM charges associated to this diagram.

%======================================================================
%======================================================================

By now, it is well understood that certain 3d CFTs with toric CY$_4$ moduli spaces can be generated by taking the same quivers and superpotentials for 4d CFTs with toric CY$_3$ moduli spaces
\cite{Martelli:2008si,Hanany:2008cd,Ueda:2008hx,Imamura:2008qs,Franco:2008um}.
The toric diagram for a CY$_3$ is 2-dimensional (more precisely, it is a plane in 3 dimensions).
The CS levels control how the parent toric diagram is ``inflated" into the 3-dimensional one for the CY$_4$.

With these ideas in mind, it is not hard to identify a candidate CS quiver for $\mathcal{C}(Q^{2,2,2})$. We just
have to consider a 4d theory whose toric diagram corresponds to collapsing the one in \fref{toric_Q222_0} onto a plane, namely
it is a square with an internal point. This is the toric diagram for a complex cone over $F_0$, i.e. a $\mathbb{Z}_2$ orbifold of the conifold \cite{Morrison:1998cs}. There are two quivers for this geometry, related in 4d by the Seiberg duality \cite{Seiberg:1994pq} (see e.g. \cite{Feng:2002zw} for details). We now check that both of them are candidates for the theory on M2-branes over $\mathcal{C}(Q^{2,2,2})$ in the sense that they give the right moduli space and chiral operator spectrum.\footnote{These models have been already considered in the context of M2-branes in \cite{Hanany:2008fj}.}

\begin{figure}[h]
\begin{center}
\includegraphics[width=9cm]{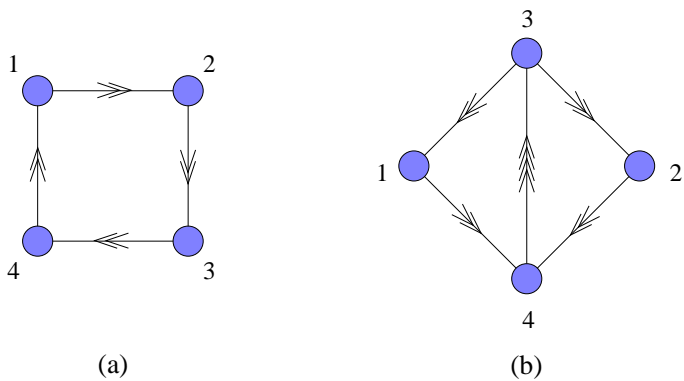}
\caption{Quiver diagrams for M2-branes over ${\cal C}(Q^{2,2,2})$. The quivers are the same as the two Seiberg dual phases for D3-branes over ${\cal C}(F_0)$.}
\label{quiver_Q222}
\end{center}
\end{figure}

Let us first consider the so called phase I. Its quiver diagram is shown in \fref{quiver_Q222}.a, and its superpotential is
\beq
W_I={\rm Tr} \, \epsilon_{ij} \epsilon_{mn} X^i_{12}X^m_{23}X^j_{34}X^n_{41} \, ,
\label{W_Q222_1}
\eeq
and the CS levels $\vec{k}=(k,k,-k,-k)$.\footnote{This quiver with a different choice of CS levels, $\vec{k}=(k,-k,k,-k)$, appeared in \cite{Benna:2008zy} as an orbifold of ABJM theory. This theory appears to describe M2-branes on $(\mathbb{C}^2/\mathbb{Z}_2)^2/\mathbb{Z}_k$ \cite{Hanany:2008cd}.} There is a $\mathbb{Z}_2$ symmetry of the theory (rotation of the quiver by 180 
degrees accompanied by the parity which flips the CS levels) that 
implies that the R-charges of the fields on the opposite sides of the quiver
are equal: $R(X^i_{12})=R(X^j_{34})$, $R(X^m_{23})=R(X^n_{41})$ (for any $i$, $j$, $m$ and $n$). The marginality of the superpotential also requires $R(X^i_{12}) + R(X^m_{23})=1$.

%======================================================================
%======================================================================

The theory has a manifest $SU(2)^2$ global symmetry, under which fields transform as
\beq
\begin{array} {c|ccc}
& SU(2)_1 & \ \ \ \ & SU(2)_2  \\ \hline
\ \ \ X_{12}^i \ \ \ & \fund & &  \\
\ \ \ X_{23}^m \ \ \ & & & \fund \\
\ \ \ X_{34}^j \ \ \ & \fund & &  \\
\ \ \ X_{41}^n \ \ \ & & & \fund
\end{array}
\eeq
As a test of the proposal, we compute the moduli space for the abelian $N=1$ gauge theory with $k=1$. We find that the moduli space is indeed ${\cal C}(Q^{2,2,2})$, whose toric diagram is shown in \fref{toric_Q222_0}. The full computation is presented in appendix \ref{appendix_Q222}, where we use the techniques in \cite{Franco:2008um}.

Following the general discussion in section \ref{section_higher_CS}, we can study the $Q^{2,2,2}$ theory at higher $k$. The action on the scalars is
\beq
X^i_{12} \sim X^i_{12}\ , \quad \quad
X^i_{34}\sim X^i_{34}\ , \quad \quad
(X^m_{23},X^m_{41})\sim (e^{i\frac{\pi}{k}}\, X^m_{23}, e^{-i
\frac{\pi}{k}}\, X^m_{41}) \, .
\eeq
We conclude that the action of the CS orbifold preserves the $SU(2)_1 \times SU(2)_2$ global symmetry.

Let us now consider the phase II quiver diagram \fref{quiver_Q222}.b; in $4$ dimensions it is related to phase I through Seiberg duality. It is interesting that the quiver for phase II corresponds to ``doubling" the one for $Q^{1,1,1}$ presented in section \ref{section_general_Q111}. The superpotential is given by
\beq
W_{II}={\rm Tr} \left(\epsilon_{ij} \, \epsilon_{mn} \, X_{32}^i X_{24}^m X_{43}^{jn} - \epsilon_{ij} \, \epsilon_{mn} \, X_{31}^m X_{14}^i X_{43}^{jn} \right) \, ,
\label{W_Q222_2}
\eeq
and the CS levels are $\vec{k}=(k,k,-k,-k)$. The theory has an $SU(2)^2$ global symmetry, under which fields transform as
\beq
\begin{array} {c|ccc}
& SU(2)_1 & \ \ \ \ & SU(2)_2  \\ \hline
\ \ \ X_{32}^i \ \ \ & \fund & &  \\
\ \ \ X_{14}^i \ \ \ & \fund & &  \\
\ \ \ X_{31}^m \ \ \ & & & \fund \\
\ \ \ X_{24}^m \ \ \ & & & \fund  \\
\ \ \ X_{43}^{im} \ \ \ & \fund & & \fund
\end{array}
\eeq
Once again, we can test the proposal by computing the moduli space for the abelian $N=1$ gauge theory with $k=1$ and verify that it is indeed ${\cal C}(Q^{2,2,2})$. The corresponding calculation is given in appendix \ref{appendix_Q222}.
For a general $k$, the scalars are identified according to
\begin{equation}
X^{im}_{43}\sim X^{im}_{43}\ ,\quad (X^i_{14},X^i_{32})\sim (e^{i
\frac{\pi}{k}}\, X^i_{14},e^{-i\frac{\pi}{k}}\, X^i_{32})\ ,\quad
(X^m_{24},X^m_{31})\sim (e^{i\frac{\pi}{k}}\, X^m_{24},e^{-i\frac{\pi}
{k}}\,  X^m_{31})\, .
\end{equation}
As before, the CS orbifold preserves the $SU(2)_1 \times SU(2)_2$ global symmetry.

%=====================================================================
\subsection{Chiral operators}
%=====================================================================

The Kaluza-Klein harmonics on $Q^{2,2,2}$ are a subset of
those on $Q^{1,1,1}$. Since the orbifold action divides the range of $\psi$ by 2, the harmonics with odd R-charge are not single-valued on
$Q^{2,2,2}$. So, before taking the $\mathbb{Z}_k$ orbifold of $Q^{2,2,2}$, we find harmonics with $SU(2)_1\times SU(2)_2\times SU(2)_3$ quantum numbers
$J_1=J_2=J_3=n$ at R-charge $2n$. The three magnetic quantum numbers $m_i$ range from $-n$ to $n$ in integer steps; thus, the total number of R-charge $2n$ states is $(2n+1)^3$.

The $\mathbb{Z}_k$ orbifold projects out some of these modes. In this case the orbifold acts by a
rotation of the third 2-sphere by $2\pi/k$ and thus breaks $SU(2)_3$. As a result, we pick out only the
$m_3=0$ states invariant under
rotations around the $z$-axis of the third $S^2$. Therefore, we are left with $(2n+1)^2$ states transforming with spin $J_1=J_2=n$ under
the remaining $SU(2)_1\times SU(2)_2$.

We now reproduce this result in the two gauge theories introduced in the previous section. Let us focus on $k\gg 1$ and consider the mesonic operators only, which do not contain monopole operators.

%===============================
\subsubsection*{Phase I}
%===============================

In this model, the construction of chiral operators is particularly simple. The analysis is exactly the same as in the 4d gauge theory dual to $AdS_5\times T^{1,1}/\mathbb{Z}_2$. We can immediately write down 16 quartic objects corresponding to all possible length 4 loops around the quiver:

\beq
X^{ij,mn}_I=X^i_{12} X^m_{23} X^j_{34} X^n_{41} \, .
\eeq
The R-charge 2 chiral operators are ${\rm Tr} X^{ij,mn}_I$, but there are only 9 of them. Applying the superpotential F-term relations to them,
\beq
X^1_{12} X^m_{23} X^2_{34} = X^2_{12} X^m_{23} X^1_{34}\ , \qquad X^1_{23} X^j_{34} X^2_{41}= X^2_{23} X^j_{34} X^1_{41}\ , \quad
etc.
\eeq
we find that the $SU(2)_1$ and $SU(2)_2$ indices are symmetrized. Therefore, these operators have $R=2$ and spins $J_1=J_2=1$. In general, the $R=2n$ chiral operators take the form
\beq
{\rm Tr} \prod_{a=1}^n X_I^{i_aj_a,m_a,n_a} \, ,
\eeq
with $SU(2)_1$ and $SU(2)_2$ indices symmetrized due to the F-term relations. These operators thus have spins $J_1=J_2=n$, matching the gravity result.

%===============================
\subsubsection*{Phase II}
%===============================

Since in $4$ dimensions this theory is a Seiberg dual of phase I, we expect to find the same spectrum of chiral operators.
Let us work it out explicitly. As a warm-up, we write down the 9 spin $(1,1)$, $R=2$, gauge-invariant chiral operators
\beq
{\rm Tr} X^i_{14} X^{jm}_{43} X^n_{31} \, ,
\eeq
where $SU(2)_1$ and $SU(2)_2$ indices are symmetrized due to the F-term equations. These operators have $R=2$ due to marginality of the superpotential \eref{W_Q222_2}. There is an additional set of operators of the same form, where we change the gauge group index $1 \to 2$. They are equal to the operators above via the F-term relation
\beq
X^i_{32} X^m_{24} = X^m_{31} X^i_{14} \,.
\eeq

In general, the $R=2n$ chiral operators are given by
\beq
{\rm Tr} \prod_{a=1}^n X_{II}^{i_aj_a,m_a,n_a} \, ,
\eeq
where $X_{II}^{ij,mn} = X^i_{14} X^{jm}_{43} X^n_{31}$.
Symmetrization over $SU(2)_1$ and $SU(2)_2$ indices follows from the superpotential F-term conditions, leading
to spin $J_1=J_2=n$ and again matches the gravity result.

%=====================================================================
\section{Chiral operators in the $M^{3,2}$ gauge theory}
%=====================================================================

\label{section_M32}

In this section we make a small digression from the main topic of this paper, namely $Q^{1,1,1}$ and its orbifolds, and study chiral operators in the gauge theory for M2-branes on ${\mathcal C}(M^{3,2})/\mathbb{Z}_k$. This theory exhibits a similar behavior to other examples we have considered: the $\mathbb{Z}_k$ orbifold preserves only the subgroup of the geometrical symmetries that is manifest in the gauge theory.

\begin{figure}[h]
\begin{center}
\includegraphics[width=4cm]{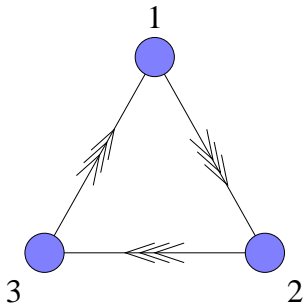}
\caption{Quiver diagram for M2-branes over $\mathcal{C}(M^{3,2})$.}
\label{quiver_M32}
\end{center}
\end{figure}

The CS gauge theory for $M^{3,2}$ was introduced in \cite{Martelli:2008si,Hanany:2008cd} and further studied in \cite{Hanany:2008fj}. The quiver diagram is shown in \fref{quiver_M32}, and the superpotential is
\beq
W={\rm Tr} \left(\epsilon_{ijk} X^i_{12} X^j_{23} X^k_{31}\right) \, , \qquad i,j,k=1\ldots 3\ . 
\label{W_M32}
\eeq 
Curiously, these are the same as in the well-known theory for D3-branes on $\mathbb{C}^3/\mathbb{Z}_3$ \cite{Kachru:1998ys,Lawrence:1998ja}.
Note that, even in the abelian theory, the superpotential does not vanish.
The CS levels are $(-2k,k,k)$. The theory has a manifest $U(1)_R\times SU(3)\times U(1)$ global symmetry, while the isometries of
$M^{3,2}$ are $U(1)_R\times SU(3)\times SU(2)$.

%======================================================================
%======================================================================

Let us first consider $N=1$ and $k=1$. In this case, the moduli space of the gauge theory was computed in \cite{Martelli:2008si,Hanany:2008cd}, and found to agree with $\mathcal{C}(M^{3,2})$. The choice of CS levels dictates that the chiral operators have to be invariant only under the $Q_2-Q_3$ combination of the $U(1)$ gauge symmetries. The simplest such operators are
\beq
X^{ijk}=X^i_{12} X^j_{23} X^k_{31}\ , \ \ \ \ \ \ \ X^{ijk}_+=X^i_{23} X^j_{31} X^k_{31}\ , \ \ \ \ \ \ \ X^{ijk}_-=X^i_{23} X^j_{12} X^k_{12} \, .
\eeq
The F-term relations
\beq
\epsilon_{ijk}\, X_{23}^j\, X_{31}^k=0\ ,\qquad \epsilon_{ijk}\,X_{31}^k\, X_{12}^i=0\ ,\qquad \epsilon_{ijk}\,X_{12}^i\, X_{23}^j=0 \, ,
\label{generators_M32}
\eeq
imply that each of the R-charge 2 operators is in the {\bf 10} of $SU(3)$, with $X^{ijk}$, $X^{ijk}_+$ and $X^{ijk}_-$ corresponding to $m=0,1,-1$ members of an $SU(2)$ triplet, respectively. This agrees with the quantum numbers of the $R=2$ spherical harmonics on
$M^{3,2}$ \cite{Fabbri:1999hw}.

Let us now turn to general $k$. Under the $\mathbb{Z}_k$ orbifold, the fields transform as
\beq
X^i_{12} \sim e^{-i\frac{\pi}{k}}\, X^i_{12}\ , \ \ \ \ \ \ X^i_{23} \sim X^i_{23}\ , \ \ \ \ \ \ X^i_{31} \sim e^{i\frac{\pi}{k}}\, X^i_{31} \, .
\eeq
This action corresponds to $e^{2\pi i J_3/k}$ and therefore breaks the $SU(2)$ part of the global symmetry.
%; it preserves just the $SU(3)\times U(1)^2$ global symmetry. 
Only the $m=0$ operators $X^{ijk}$ in \eref{generators_M32} are invariant. 

In the non-abelian theory, the single-trace gauge invariant chiral operators assume the form
\beq
{\rm Tr} \prod_{a=1}^n X^{i_a j_a k_a} \, .
\eeq
These operators have R-charge $R=2n$ and are in the symmetric $3n$-box representations of $SU(3)$ due to F-term relations. Let us compare this with the spectrum of spherical harmonics. For $M^{3,2}$ one finds that hypermultiplet states with
$R=2n$ are in the symmetric $3n$-box representations of $SU(3)$, and in the spin $J=n$ representation of $SU(2)$ \cite{Fabbri:1999hw}. 
For $M^{3,2}/\mathbb{Z}_k$ the $SU(2)$ is broken by the action $e^{2\pi i J_3/k}$, and we must retain only the $m=0$ state from each $SU(2)$ multiplet.
The resulting spectrum agrees with the gauge invariant operators we have constructed.

%=====================================================================
\section{Resolutions of ${\mathcal C}(Q^{1,1,1})$}
%=====================================================================

\label{section_resolutions}

\begin{figure}[h]
\begin{center}
\includegraphics[width=10cm]{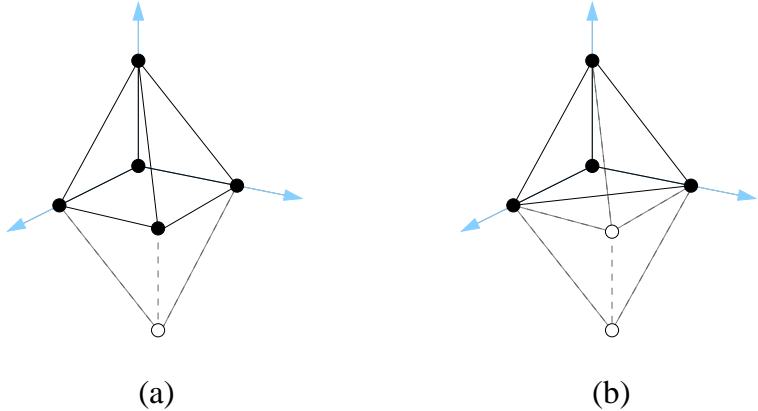}
\caption{Resolutions of $\mathcal{C}(Q^{1,1,1})$. Starting from the toric diagram in \fref{toric_quiver_Q111}.a we remove points (indicated with white circles). This operation results in: a) $\mathcal{C}(T^{1,1})\times \mathbb{C}$ and b) $\mathbb{C}^4$.}
\label{toric_resolutions_Q111}
\end{center}
\end{figure}

In this section, we investigate possible symmetry breaking states in the $Q^{1,1,1}$ candidate theory \cite{Franco:2008um}. Experience with D3-branes on the conifold \cite{Klebanov:1999tb} suggests that their dual gravity description is expected to involve M2-branes on {\it resolved} cones over $Q^{1,1,1}$.
$\mathcal{C}(Q^{1,1,1})$ is a $\mathbb{C}^2$ bundle over $\mathbb{P}^1 \times \mathbb{P}^1$; its resolutions correspond to blowing-up the $\mathbb{P}^1$'s. Blowing-up one $\mathbb{P}^1$ produces $\mathcal{C}(T^{1,1})\times \mathbb{C}$. A generic blow-up of the remaining $\mathbb{P}^1$ resolves the singularity completely, resulting in $\mathbb{C}^4$. The sequence of resolutions is then
\begin{equation}
\mathcal{C}(Q^{1,1,1}) \ \ \ \rightarrow \ \ \ \mathcal{C}(T^{1,1})\times \mathbb{C} \ \ \ \rightarrow \ \ \ \mathbb{C}^4\, .
\label{resolution_sequence_Q111}
\end{equation}
This sequence is nicely described in terms of toric diagrams as shown in \fref{toric_resolutions_Q111}. In this language, blowing-up a $\mathbb{P}^1$ corresponds to removing a point. For higher $k$ the resulting space will be sensitive to the orientation between the blown-up $\mathbb{P}^1$ and the orbifolded ones. For a single blown-up $\mathbb{P}^1$ we should then expect two possibilities depending whether it is orbifolded or not.

%======================================================================
%======================================================================%

From a field theory perspective, resolutions correspond to turning on VEV's for the scalar component
of a chiral superfield. These VEV's break conformal invariance. Flowing to energies much lower
than the scale set by the VEV's, we obtain a new CFT that results from Higgsing
gauge groups and integrating out massive fields. In the theories we are considering, the gauge group is $U(N)
$. Thus, an FI term is required to achieve the resolutions. Such
supersymmetric FI deformations have been studied in \cite{Gomis:2008vc} and recently considered for resolution purposes in
\cite{Gaiotto:2009tk}.

In the next section, we restrict to the abelian theory and compute the moduli space of the resulting IR CFT after turning on VEV's. We then compare this geometry with the one resulting from blowing-up $\mathbb{P}^1$'s, finding agreement. This matching provides further support for our
identification of the $Q^{1,1,1}$ theory (and its orbifolds).

It is important to emphasize that, although the abelian intuition provides valuable guidance in the determination of new theories, it does not probe their non-abelian structure. Thus the theories we obtain by turning on VEV's should be regarded as potential candidates for new M2-brane theories, but further checks are required to determine whether they can be promoted to non-abelian theories on stacks of M2-branes.

%=====================================================================
\subsection{Symmetry breaking in the gauge theory}
%=====================================================================

\label{section_resolution_FT}

We presented the $Q^{1,1,1}$ quiver in \fref{toric_quiver_Q111}.b and its superpotential in \eref{W_Q111}. The are two distinct options for blow-ups: either giving a VEV to one of the internal fields (namely to one $A_i$) or to one of the external ones (a $B_i$ or a $C_i$). We now investigate the two alternatives.

\bigskip

\begin{figure}[h]
\begin{center}
\includegraphics[width=4.5cm]{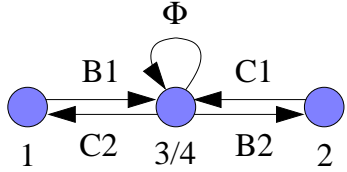}
\caption{Quiver diagram for a partial resolution of the $Q^{1,1,1}$ theory. For the abelian theory, the moduli space is $\mathbb{C}\times \mathcal{C}(T^{1,1})$.}
\label{fig3}
\end{center}
\end{figure}

\noindent{\bf a) Turning on a VEV for $A_1$:} the quiver becomes that in \fref{fig3} where we have renamed $A_2=\Phi$, and the superpotential is
\begin{equation}
W=\Phi\Big(C_2B_1B_2C_1-B_2C_1C_2B_1\Big)\, .
\label{W_resolution_1}
\end{equation}

%======================================================================
%======================================================================%

Nodes 3 and 4 are combined into one node which we indicate as $3/4$; this corresponds to breaking of $U(N)_3\times U(N)_4$ to the diagonal $U(N)$ subgroup. Starting from $Q^{1,1,1}$ with $\vec{k}=(1,1,-1,-1)$, we end up with $(k_1,k_2,k_{3/4})=(1,1,-2)$ (the CS levels of the higgsed gauge groups are added). We can choose the effective D-terms to be given by the combination $Q_2-Q_1$. The resulting invariants are
\beq
\begin{array}{lclcl}
z_1=B_1 C_2 & \ \ \ \ \ & z_2=B_2 C_1 & \ \ \ \ \ & w=\Phi \\
z_3=B_1 C_1 & \ \ \ \ \ & z_4=B_2 C_2 & \ \ \ \ \ &
\end{array}
\eeq
As might have been expected, the adjoint field parameterizes a $\mathbb{C}$ factor, while the $z_i$ (made out of $B_i$ and $C_i$) parameterize a conifold.
In fact, even though it is not a necessary condition, the superpotential \eref{W_resolution_1} factorizes as the adjoint times the conifold superpotential. We see that the gauge theory computation reproduces the geometric expectation when blowing up a $\mathbb{P}^1$.

Let us now consider the general $k$ case. In the IR, we now have $(k_1,k_2,k_{3/4})=(k,k,-2k)$. After fixing the gauge, we are left with the following discrete identifications
\beq
\begin{array}{lcl}
B_1\sim B_1 \, e^{i\frac{\pi}{k}} & \ \ \ \ \ & B_2\sim B_2 \, e^{-i\frac{\pi}{k}} \\
C_1\sim C_1 \, e^{i\frac{\pi}{k}}  & \ \ \ \ \ & C_2\sim C_2 \, e^{-i\frac{\pi}{k}}
\end{array}
\eeq
This translates into
\beq
z_3\sim z_3 \, e^{i\frac{2\pi}{k}} \ \ \ \ \ z_4\sim z_4 \, e^{-i\frac{2\pi}{k}}\, ,
\eeq
without any identification for the $\mathbb{C}$ factor.

\bigskip
\bigskip

\begin{figure}[h]
\begin{center}
\includegraphics[width=3.5cm]{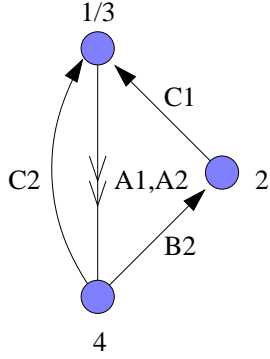}
\caption{Quiver diagram for another partial resolution of the $Q^{1,1,1}$ theory. For the abelian theory, the moduli space is $\mathbb{C}\times \mathcal{C}(T^{1,1})$.}
\label{fig2}
\end{center}
\end{figure}

%======================================================================%
\noindent{\bf b) Turning on a VEV for $B_1$:} the quiver becomes
the one in \fref{fig2}, with superpotential
\begin{equation}
W=C_2 \, A_1\, B_2\, C_1\, A_2-C_2\, A_2\, B_2\, C_1\, A_1\, .
\end{equation}

%======================================================================
%======================================================================
%
Nodes 1 and 3 are now combined into one; this corresponds to breaking of $U(N)_1\times U(N)_3$ to the diagonal $U(N)$ subgroup. Starting from $Q^{1,1,1}$ with $\vec{k}=(1,1,-1,-1)$ we are left with $(k_{1/3},k_2,k_4)=(0,1,-1)$.
We can take $Q_2+Q_4$ to give the effective D-terms. The resulting invariants are
\beq
\begin{array}{lclcl}
z_1=A_1 C_1 & \ \ \ \ \ & z_2=A_2 C_2 & \ \ \ \ \ & w=B_2 \\
z_3=A_2 C_1 & \ \ \ \ \ & z_4=A_1 C_2 & \ \ \ \ \ &
\end{array}
\eeq
Clearly $z_1z_2-z_3z_4=0$, so the $z_i$ parametrize $\mathcal{C}(T^{1,1})$ while $w$ parametrizes $\mathbb{C}$. The moduli space is once again $\mathbb{C}\times \mathcal{C}(T^{1,1})$, in agreement with the geometric expectation.

Let us now take general $k$. In the IR, we are left with $(k_{1/3},k_2,k_4)=(0,k,-k)$. Repeating the computation above, we obtain the discrete identifications
\beq
\begin{array}{lclcl}
A_1\sim A_1 \, e^{i\frac{\pi}{k}} & \ \ \ \ \ & A_2\sim A_2 \, e^{i\frac{\pi}{k}} & \ \ \ \ \ &  B_2\sim B_2 \, e^{-i\frac{2\pi}{k}} \\
C_1\sim C_1 \, e^{i\frac{\pi}{k}} & \ \ \ \ \ & C_2\sim C_2 \, e^{-i\frac{\pi}{k}}  & \ \ \ \ \ &
\end{array}
\eeq
which translate into
\beq
z_1\sim z_1e^{i\frac{2\pi}{k}} \ \ \ \ \ z_3\sim z_3 e^{i\frac{2\pi}{k}} \ \ \ \ \ w\sim we^{-i\frac{2\pi}{k}}\, .
\eeq
This is a somewhat different $\mathbb{Z}_k$ orbifold of $\mathbb{C}\times \mathcal{C}(T^{1,1})$.

\bigskip

%======================================================================
\centerline{\bf Acknowledgements} \vskip 5mm \noindent
%======================================================================

S.F. and D. R-G, would like to thank A. Hanany and J. Park for previous collaboration on a related project \cite{Franco:2008um}. I.R.K. is grateful to M. Benna, T. Klose, A. Murugan and M. Smedb\" ack for discussions and collaboration on related topics. S.F. is supported by the DOE under contract DE-FG02-91ER-40671 and by the National Science Foundation under Grant No. PHY05-51164. I.R.K. is supported by the National Science Foundation under Grant No. PHY-0756966. D. R-G. acknowledges financial support from the European Commission through Marie Curie OIF grant contract no. MOIF-CT- 2006-38381.

%%%%%%%%%%%%%%%%%%%%%%%%%%%%%%%%%%%%%%%%%%%%%%%%%%%%%%%%%%%%%%%%%%%%
\appendix

%=====================================================================
\section{Moduli space of the $\mathbb{Z}_2$ orbifold of the quiver}
%=====================================================================

\label{appendix_Z2_Z2}

Here we use the techniques of \cite{Franco:2008um} to compute the moduli
space of the theory introduced in section \ref{section_Q111_Z2_Z2} in
the abelian, $N=1$ case with $k=1$. We find a new example of a
phenomenon already observed for orbifolds of the ABJM theory: a
$\mathbb{Z}_p$ orbifold projection of the quiver \cite{Benna:2008zy}
leads to a $\mathbb{Z}_p \times \mathbb{Z}_p$ orbifold of the moduli
space \cite{Imamura:2008nn,Terashima:2008ba}.

The quiver and GLSM fields are related by the matrix

{\scriptsize
\beq
P=\left(\begin{array}{c|cccccccccccccccccccc}
\ \ \ \ \ & \ p_1 \ & \ p_2 \ & \ p_3 \ & \ p_4 \ & \ p_5 \ & \ p_6 \ & \
p_7 \ & \ p_8 \ & \ p_9 \ & \ p_{10} \ & \ p_{11} \ & \ p_{12} \ & \ p_{13} \ & \
p_{14} \ & \ p_{15} \ & \
p_{16} \ & \ p_{17} \ & \ p_{18} \ & \ p_{19} \ & \ p_{20} \ \\
\hline
A^1_1 & 1 & 1 & 0 & 0 & 0 & 0 & 0 & 0 & 0 & 0 & 0 & 0 & 0 & 0 & 0 & 0 & 0 & 0 & 0 & 0 \\
A^2_1 & 0 & 1 & 0 & 1 & 0 & 0 & 0 & 0 & 0 & 0 & 0 & 0 & 0 & 0 & 0 & 0 & 0 & 0 & 0 & 0 \\
A^1_2 & 1 & 0 & 1 & 0 & 0 & 0 & 0 & 0 & 0 & 0 & 0 & 0 & 0 & 0 & 0 & 0 & 0 & 0 & 0 & 0 \\
A^2_2 & 0 & 0 & 1 & 1 & 0 & 0 & 0 & 0 & 0 & 0 & 0 & 0 & 0 & 0 & 0 & 0 & 0 & 0 & 0 & 0 \\
B^1_1 & 0 & 0 & 0 & 0 & 1 & 1 & 0 & 0 & 0 & 0 & 0 & 0 & 1 & 0 & 0 & 0 & 1 & 0 & 0 & 0 \\
B^2_1 & 0 & 0 & 0 & 0 & 0 & 1 & 0 & 1 & 0 & 1 & 0 & 1 & 0 & 0 & 0 & 0 & 0 & 0 & 0 & 0 \\
B^1_2 & 0 & 0 & 0 & 0 & 1 & 0 & 1 & 0 & 1 & 0 & 1 & 0 & 0 & 0 & 0 & 0 & 0 & 0 & 0 & 0 \\
B^2_2 & 0 & 0 & 0 & 0 & 0 & 0 & 1 & 1 & 0 & 0 & 0 & 0 & 0 & 1 & 0 & 0 & 0 & 1 & 0 & 0 \\
C^1_1 & 0 & 0 & 0 & 0 & 0 & 0 & 0 & 0 & 1 & 1 & 0 & 0 & 0 & 0 & 1 & 0 & 0 & 0 & 1 & 0 \\
C^2_1 & 0 & 0 & 0 & 0 & 0 & 0 & 0 & 0 & 0 & 0 & 0 & 0 & 1 & 1 & 1 & 1 & 0 & 0 & 0 & 0 \\
C^1_2 & 0 & 0 & 0 & 0 & 0 & 0 & 0 & 0 & 0 & 0 & 1 & 1 & 0 & 0 & 0 & 1 & 0 & 0 & 0 & 1 \\
C^2_2 & 0 & 0 & 0 & 0 & 0 & 0 & 0 & 0 & 0 & 0 & 0 & 0 & 0 & 0 & 0 & 0 & 1 & 1 & 1 & 1
\end{array}\right)
\label{P_Q111_Z2_Z2}
\eeq}
The charge matrix for the F-term constraints is then

{\scriptsize
\beq
Q_F=\left(\begin{array}{cccccccccccccccccccc}
\ p_1 \ & \ p_2 \ & \ p_3 \ & \ p_4 \ & \ p_5 \ & \ p_6 \ & \
p_7 \ & \ p_8 \ & \ p_9 \ & \ p_{10} \ & \ p_{11} \ & \ p_{12} \ & \ p_{13} \ & \
p_{14} \ & \ p_{15} \ & \
p_{16} \ & \ p_{17} \ & \ p_{18} \ & \ p_{19} \ & \ p_{20} \ \\
\hline
0 & 0 & 0 & 0 & 1 & 0 & 0 & 0 & 0 & 0 & -1 & 0 & 0 & 0 & 0 & 0 & -1 & 0 & 0 & 1 \\
0 & 0 & 0 & 0 & 1 & 0 & 0 & 0 & -1 & 0 & 0 & 0 & 0 & 0 & 0 & 0 & -1 & 0 & 1 & 0 \\
0 & 0 & 0 & 0 & 1 & 0 & -1 & 0 & 0 & 0 & 0 & 0 & 0 & 0 & 0 & 0 & -1 & 1 & 0 & 0 \\
0 & 0 & 0 & 0 & 1 & 0 & 0 & 0 & 0 & 0 & -1 & 0 & -1 & 0 & 0 & 1 & 0 & 0 & 0 & 0 \\
0 & 0 & 0 & 0 & 1 & 0 & 0 & 0 & -1 & 0 & 0 & 0 & -1 & 0 & 1 & 0 & 0 & 0 & 0 & 0 \\
0 & 0 & 0 & 0 & 1 & 0 & -1 & 0 & 0 & 0 & 0 & 0 & -1 & 1 & 0 & 0 & 0 & 0 & 0 & 0 \\
0 & 0 & 0 & 0 & 1 & -1 & 0 & 0 & 0 & 0 & -1 & 1 & 0 & 0 & 0 & 0 & 0 & 0 & 0 & 0 \\
0 & 0 & 0 & 0 & 1 & -1 & 0 & 0 & -1 & 1 & 0 & 0 & 0 & 0 & 0 & 0 & 0 & 0 & 0 & 0 \\
0 & 0 & 0 & 0 & 1 & -1 & -1 & 1 & 0 & 0 & 0 & 0 & 0 & 0 & 0 & 0 & 0 & 0 & 0 & 0 \\
1 & -1 & -1 & 1 & 0 & 0 & 0 & 0 & 0 & 0 & 0 & 0 & 0 & 0 & 0 & 0 & 0 & 0 & 0 & 0
\end{array}\right)
\label{QF_Q111_Z2_Z2}
\eeq}

The quiver charges are given by

{\scriptsize
\beq
\begin{array}{c|cccccccccccccccccccc}
\ \ \ \ \ & \ p_1 \ & \ p_2 \ & \ p_3 \ & \ p_4 \ & \ p_5 \ & \ p_6 \ & \
p_7 \ & \ p_8 \ & \ p_9 \ & \ p_{10} \ & \ p_{11} \ & \ p_{12} \ & \ p_{13} \ & \
p_{14} \ & \ p_{15} \ & \
p_{16} \ & \ p_{17} \ & \ p_{18} \ & \ p_{19} \ & \ p_{20} \ \\
\hline

\ \ Q^1_1 \ \ & 0 & 0 & 0 & 0 & -1 & 0 & 0 & 0 & 0 & 0 & 1 & 0 & 0 & 0 & 0 & 0 & 0 & 0 & 0 & 0 \\
\ \ Q^2_1 \ \ & 0 & 0 & 0 & 0 & 0 & 0 & 0 & -1 & 0 & 0 & 0 & 0 & 0 & 0 & 0 & 0 & 0 & 1 & 0 & 0 \\
\ \ Q^1_2 \ \ & 0 & 0 & 0 & 0 & 0 & 0 & 1 & 0 & -1 & 0 & 0 & 0 & 0 & 0 & 0 & 0 & 0 & 0 & 0 & 0 \\
\ \ Q^2_2 \ \ & 0 & 0 & 0 & 0 & 1 & 0 & 0 & 0 & 0 & 0 & 0 & 0 & -1 & 0 & 0 & 0 & 0 & 0 & 0 & 0 \\
\ \ Q^1_3 \ \ & -1 & 0 & 0 & 0 & 0 & 0 & 0 & 0 & 0 & 1 & 0 & 0 & 0 & 0 & 0 & 0 & 0 & 0 & 0 & 0 \\
\ \ Q^2_3 \ \ & 0 & 0 & 0 & -1 & 0 & 0 & 0 & 0 & 0 & 0 & 0 & 0 & 1 & 0 & 0 & 0 & 0 & 0 & 0 & 0 \\
\ \ Q^1_4 \ \ & 1 & 0 & 0 & 0 & 0 & 0 & 0 & 0 & 0 & 0 & -1 & 0 & 0 & 0 & 0 & 0 & 0 & 0 & 0 & 0 \\
\ \ Q^2_4 \ \ & 0 & 0 & 0 & 1 & 0 & 0 & -1 & 1 & 1 & -1 & 0 & 0 & 0 & 0 & 0 & 0 & 0 & -1 & 0 & 0
\end{array}
\label{Q_Q111_Z2_Z2}
\eeq}

The CS levels are $k=(1,1,1,1,-1,-1,-1,-1)$, where the order of nodes is $(V^1_1,V^2_1,V^1_2,V^2_2,V^1_3,V^2_3,V^1_4,V^2_4)$. Then, we can take $(Q^1_1+Q^1_3,Q^1_1+Q^2_3,Q^1_1+Q^1_4,Q^1_1+Q^2_4,Q^2_1+Q^1_3,Q^1_2+Q^1_3)$ as D-terms,

{\scriptsize
\beq
Q_D=\left( \begin{array}{cccccccccccccccccccc}
\ p_1 \ & \ p_2 \ & \ p_3 \ & \ p_4 \ & \ p_5 \ & \ p_6 \ & \
p_7 \ & \ p_8 \ & \ p_9 \ & \ p_{10} \ & \ p_{11} \ & \ p_{12} \ & \ p_{13} \ & \
p_{14} \ & \ p_{15} \ & \
p_{16} \ & \ p_{17} \ & \ p_{18} \ & \ p_{19} \ & \ p_{20} \ \\
\hline
-1 & 0 & 0 & 0 & -1 & 0 & 0 & 0 & 0 & 1 & 1 & 0 & 0 & 0 & 0 & 0 & 0 & 0 & 0 & 0 \\
0 & 0 & 0 & -1 & -1 & 0 & 0 & 0 & 0 & 0 & 1 & 0 & 1 & 0 & 0 & 0 & 0 & 0 & 0 & 0 \\
1 & 0 & 0 & 0 & -1 & 0 & 0 & 0 & 0 & 0 & 0 & 0 & 0 & 0 & 0 & 0 & 0 & 0 & 0 & 0 \\
0 & 0 & 0 & 1 & -1 & 0 & -1 & 1 & 1 & -1 & 1 & 0 & 0 & 0 & 0 & 0 & 0 & -1 & 0 & 0 \\
-1 & 0 & 0 & 0 & 0 & 0 & 0 & -1 & 0 & 1 & 0 & 0 & 0 & 0 & 0 & 0 & 0 & 1 & 0 & 0 \\
-1 & 0 & 0 & 0 & 0 & 0 & 1 & 0 & -1 & 1 & 0 & 0 & 0 & 0 & 0 & 0 & 0 & 0 & 0 & 0
\end{array}
\right)
\label{QD_Q111_Z2_Z2}
\eeq}

\begin{figure}[h]
\begin{center}
\includegraphics[width=6cm]{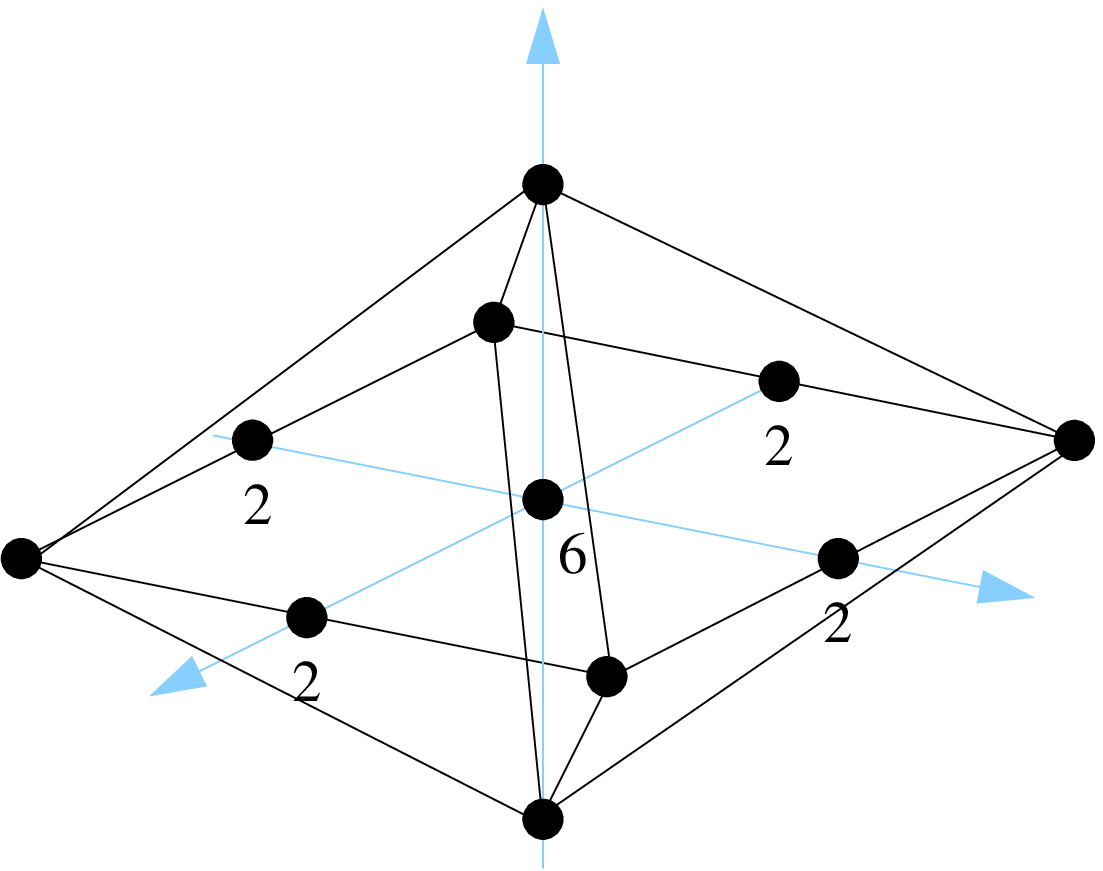}
\caption{Toric diagram for the moduli space of the theory introduced in section \ref{section_Q111_Z2_Z2}. It corresponds to ${\cal C}(Q^{1,1,1}/(\mathbb{Z}_2 \times \mathbb{Z}_2))$. The numbers indicate the multiplicity of the corresponding GLSM fields.}
\label{toric_Q111_Z2_Z2}
\end{center}
\end{figure}

The toric diagram is computed as the kernel of $Q_{tot}=(Q_F,Q_D)$, and is given
by
{\scriptsize
\beq
G^T=\left( \begin{array}{cccccccccccccccccccc}
\ p_1 \ & \ p_2 \ & \ p_3 \ & \ p_4 \ & \ p_5 \ & \ p_6 \ & \
p_7 \ & \ p_8 \ & \ p_9 \ & \ p_{10} \ & \ p_{11} \ & \ p_{12} \ & \ p_{13} \ & \
p_{14} \ & \ p_{15} \ & \
p_{16} \ & \ p_{17} \ & \ p_{18} \ & \ p_{19} \ & \ p_{20} \ \\
\hline
0 & 0 & 0 & 0 & 0 & 1 & -1 & 0 & -1 & 0 & 0 & 1 & 0 & -1 & -1 & 0 & 1 & 0 & 0 & 1 \\
1 & 2 & 0 & 1 & 1 & 2 & 0 & 1 & 1 & 2 & 0 & 1 & 2 & 1 & 2 & 1 & 1 & 0 & 1 & 0 \\
0 & 0 & 0 & 0 & 0 & -2 & 2 & 0 & 1 & -1 & 1 & -1 & -1 & 1 & 0 & 0 & -1 & 1 & 0 & 0 \\
0 & -1 & 1 & 0 & 0 & 0 & 0 & 0 & 0 & 0 & 0 & 0 & 0 & 0 & 0 & 0 & 0 & 0 & 0 & 0
\end{array}
\right)
\label{G_Q111_Z2_Z2}
\eeq}
All columns add up to one. We can drop, for example, the second row. Applying an $SL(3,\mathbb{Z})$ transformation, we take the toric diagram to the simple form in \fref{toric_Q111_Z2_Z2}. The toric diagram of $\mathcal{C}(Q^{1,1,1})$ is refined by a factor 2 in two directions, hence the moduli space is $\mathcal{C}(Q^{1,1,1}/(\mathbb{Z}_2 \times \mathbb{Z}_2))$.

%======================================================================
%======================================================================

%=====================================================================
\subsection*{Moduli space at CS level $k$}
%=====================================================================

Let us now study the abelian $N=1$ theory at general $k$. Since we focus on the abelian case, and to simplify comparison, we drop any ordering of fields associated with the quiver in all expressions that follow.

Let us first consider $k=1$. The $\mathbb{Z}_2\times\mathbb{Z}_2$ orbifold action on $\mathcal{C}(Q^{1,1,1})$ is

{\small
\beq
\begin{array}{ccccc}
\mathbb{Z}_2|_1\,: & \quad & (w_1,w_2,w_3,w_4,w_5,w_6,w_7,w_8) & \rightarrow & (-w_1,-w_2,-w_3,-w_4,w_5,w_6,w_7,w_8) \\
\mathbb{Z}_2|_2\,: & \quad & (w_1,w_2,w_3,w_4,w_5,w_6,w_7,w_8) & \rightarrow &  (w_1,w_2,w_3,w_4,-w_5,-w_6,-w_7,-w_8)
\end{array}
\eeq
}
where the $\{w_i\}$ are the complex variables in $\mathcal{C}(Q^{1,1,1})$. The next step is to construct the monomials that are invariant under the orbifold action, which take the general form $z_a=w_iw_j$. Taking into account the $Q^{1,1,1}$ relations satisfied by the $w_i$, we are left with 15 independent monomials. Modding by the equivalence relations of the underlying $Q^{1,1,1}$ (which we collectively denote by $\mathbb{I}(Q^{1,1,1})$), we have that the coordinate ring of the variety is $\mathbb{C}[z_a]/\mathbb{I}(Q^{1,1,1})$. More explicitly

\begin{equation}
\frac{\mathbb{C}
[w_1^2,w_2^2,w_3^2,w_4^2,w_1w_2,w_1w_3,w_1w_4,w_2w_3,w_2w_4,w_5^2,w_6^2,w_5w_6,w_7^2,w_8^2,w_7w_8]}
{\mathbb{I}(Q^{1,1,1})} \, .
\end{equation}

Let us now turn to gauge theory. The operators invariant under the $U(1)$ actions defined by \eref{QD_Q111_Z2_Z2} are

\beq
\begin{array}{ccccccc}
z_1 &=& A^1_1\, A^2_1\, B^1_2\, B^2_2\, C^1_1\, C^2_1 & \ \ \ \ \ \ &
z_2 &=&  A^1_2\, A^2_2\, B^1_1\, B^2_1\, C^1_2\, C^2_2 \\
z_3 &=& A^1_1\, A^2_1\, B^1_1\, B^2_1\, C^1_2\, C^2_2& \ \ \ \ \ \ &
z_4 &=& A^1_2\, A^2_2\, B^1_2\, B^2_2\, C^1_1\, C^2_1 \\
z_5 &=& A^1_1\, A^2_2\, B^1_1\, B^2_2\, C^1_1\, C^1_2 & \ \ \ \ \ \ &
z_6 &=& A^1_1\, A^2_1\, B^1_1\, B^2_2\, C^1_1\, C^1_2 \\
z_7 &=& A^1_1\, A^2_2\, B^1_2\, B^2_2\, C^1_1\, C^2_1  & \ \ \ \ \ \ &
z_8 &=& A^1_1\, A^2_2\, B^1_1\, B^2_1\, C^1_2\, C^2_2  \\
z_9 &=& A^1_2\, A^2_2\, B^1_1\, B^2_2\, C^1_1\, C^1_2  & \ \ \ \ \ \ &
z_{10} &=& A^1_1\, A^2_1\, B^1_1\, B^2_1\, C^1_1\, C^2_1   \\
z_{11} &=& A^1_2\, A^2_2\, B^1_1\, B^2_1\, C^1_1\, C^2_1 & \ \ \ \ \ \ &
z_{12} &=& A^1_1\, A^2_2\, B^1_1\, B^2_1\, C^1_1\, C^2_1  \\
z_{13} &=& A^1_1\, A^2_1\, B^1_2\, B^2_2\, C^1_2\, C^2_2  & \ \ \ \ \ \ &
z_{14} &=& A^1_2\, A^2_2\, B^1_2\, B^2_2\, C^1_2\, C^2_2  \\
z_{15} &=& A^1_1\, A^2_2\, B^1_2\, B^2_2\, C^1_2\, C^2_2 & \ \ \ \ \ \ &
\end{array}
\eeq
Notice that while $z_1$ to $z_9$ are invariant under the full gauge symmetry of the quiver, $z_{10}$ to $z_{15}$ require monopole operators. One can verify that these operators are in one to one correspondence with the $z_a$ and they satisfy the same relations. We thus conclude, from a gauge theory calculation alternative to the one in the previous section, that the moduli space of the theory is $\mathcal{C}(Q^{1,1,1}/(\mathbb{Z}_2\times \mathbb{Z}_2))$.

Let us now consider general $k$. The $\mathbb{Z}_k$ orbifold acts on the chiral operators as
{\small
\beq
\begin{array}{l}
(z_1,z_2,z_3,z_4,z_5,z_6,z_7,z_8,z_9,z_{10},z_{11},z_{12},z_{13},z_{14},z_{15})\rightarrow\\
(z_1,z_2,z_3,z_4,z_5,z_6,z_7,z_8,z_9,e^{i\frac{2\pi}{k}}z_{10},e^{i\frac{2\pi}{k}}z_{11},z^{i\frac{2\pi}{k}}z_{12},e^{-i\frac{2\pi}{k}}z_{13},e^{-i\frac{2\pi}{k}}z_{14},e^{-i\frac{2\pi}{k}}z_{15})
\end{array}
\eeq}
The orbifold acts on $\{z_{10},z_{11},z_{12},z_{13},z_{14},z_{15}\}$, which in terms of the original $Q^{1,1,1}$ coordinates is the set $\{w_5,w_6,w_7,w_8\}$. We thus conclude that the moduli space at higher general $k$ is

\begin{equation}
\mathcal{C}\left(\frac{Q^{1,1,1}}{\mathbb{Z}_2\times \mathbb{Z}_{2k}}\right) \,.
\end{equation}

%=====================================================================
\section{Moduli space of the $Q^{2,2,2}$ theories}
%=====================================================================

\label{appendix_Q222}

Let us compute the moduli spaces for the abelian $N=1$ case of the two theories in section \ref{section_Q222} with $k=1$.

%=================================
\subsubsection*{Phase I}
%=================================

Quiver and GLSM fields are related by

{\footnotesize
\beq
P=\left(\begin{array}{c|c c c c c c c c}
\ \ \ \ \ & \ \ p_1 \ \ & \ \ p_2 \ \ & \ \ p_3 \ \ & \ \ p_4 \ \ & \
\ p_5 \ \ & \ \ p_6 \ \ & \ \
p_7 \ \ & \ \ p_8 \\ \hline
X^1_{12} & 1 & 1 & 0 & 0 & 0 & 0 & 0 & 0\\
X^2_{12} & 1 & 0 & 0 & 0 & 1 & 0 & 0 & 0\\
X^1_{23} & 0 & 0 & 1 & 1 & 0 & 0 & 0 & 0\\
X^2_{23} & 0 & 0 & 1 & 0 & 0 & 0 & 1 & 0\\
X^1_{34} & 0 & 1 & 0 & 0 & 0& 1 &0 &0\\
X^2_{34} & 0 & 0 & 0 & 0 & 1& 1 &0 &0\\
X^1_{41} & 0 & 0 & 0 & 1 & 0& 0 &0 &1\\
X^2_{41} & 0 & 0 & 0 & 0 & 0& 0 &1 &1\\
\end{array}\right)
\eeq
}
Then, F-terms are implemented by the matrix

{\footnotesize
\beq
Q_F=\left( \begin{array}{c c c c c c c c}
\ \ p_1 \ \ & \ \ p_2 \ \ & \ \ p_3 \ \ & \ \ p_4 \ \ & \
\ p_5 \ \ & \ \ p_6 \ \ & \ \
p_7 \ \ & \ \ p_8 \\ \hline
0 & 0 & 1 & -1 & 0 & 0 & -1 & 1\\
1 &-1 & 0 & 0 & -1 & 1 & 0 &0
\end{array}\right)
\eeq
}

The quiver charges associated with GLSM fields are

{\footnotesize
\beq
\begin{array}{c | c c c c c c c c}
\ \ \ \ \ & \ \ p_1 \ \ & \ \ p_2 \ \ & \ \ p_3 \ \ & \ \ p_4 \ \ & \
\ p_5 \ \ & \ \ p_6 \ \ & \ \
p_7 \ \ & \ \ p_8 \\ \hline
\ \ Q_1 \ \ & -1 & 0 & 0 & 0 & 0 & 0 & 0 & 1\\
\ \ Q_2 \ \ & 1 & 0 & -1 & 0 & 0 & 0 & 0 & 0\\
\ \ Q_3 \ \ & 0 & 0 & 1 & 0 & 0 & -1 & 0 & 0\\
\ \ Q_4 \ \ & 0 & 0 & 0 & 0 & 0 & 1 & 0 & -1\\
\end{array}
\eeq
}

We consider CS levels $\vec{k}=(1,1,-1,-1)$. Hence, we can take effective D-terms given by the combinations $Q_1-Q_2$ and $Q_1+Q_3$.

{\footnotesize
\beq
Q_D=\left(\begin{array}{cc c c c c c c c}
\ \ p_1 \ \ & \ \ p_2 \ \ & \ \ p_3 \ \ & \ \ p_4 \ \ & \
\ p_5 \ \ & \ \ p_6 \ \ & \ \ p_7 \ \ & \ \ p_8 \\ \hline
-2 & 0 & 1 & 0 & 0 & 0 & 0 & 1 \\
-1 & 0 & 1 & 0 & 0 & -1 & 0 & 1 \\
\end{array} \right)
\eeq
}
The toric diagram is finally given by

{\footnotesize
\beq
G^T=\left(\begin{array}{c c c c c c c c}
\ \ p_1 \ \ & \ \ p_2 \ \ & \ \ p_3 \ \ & \ \ p_4 \ \ & \
\ p_5 \ \ & \ \ p_6 \ \ & \ \ p_7 \ \ & \ \ p_8 \\ \hline
 0 & 0 & -1 & 0 & 0 & 0 &0 & 1 \\
 0 & 0 & 0 & -1 & 0 & 0 & 1 & 0 \\
 1 & 2 & 2 & 2 & 0 & 1 & 0 & 0\\
 0 & -1 & 0 & 0 & 1 & 0 & 0 & 0
 \end{array}\right)
\eeq
}
All columns add up to one. Dropping the third row, we have the toric diagram for $\mathcal{C}(Q^{2,2,2})$ shown in \fref{toric_Q222_0}, with multiplicity 2 for the GLSM fields in the node at the center.

Simple inspection of $Q_{tot}$, indicates that we indeed have an additional $SU(2)_3$ symmetry. The GLSM fields transform according

{\footnotesize
\beq
\begin{array} {c|ccccc}
& SU(2)_1 & \ \ \ \ & SU(2)_2 & \ \ \ \ & SU(2)_3 \\ \hline
\ \ \ (p_2,p_5) \ \ \ & \fund & & & &  \\
\ \ \ (p_4,p_7) \ \ \ & & & \fund & &  \\
\ \ \ (p_3,p_8) \ \ \ & & & & & \fund
\end{array}
\label{SU2_p}
\eeq}
and the rest are singlets, i.e. each $SU(2)$ factor exchanges the GLSM fields on opposite corners of the toric diagram.

%=================================
\subsubsection*{Phase II}
%=================================

The matrix relating the quiver and GLSM fields is

{\footnotesize
\beq
P=\left(\begin{array}{c|ccccccccc}
\ \ \ \ \ & \ \ p_1 \ \ & \ \ p_2 \ \ & \ \ p_3 \ \ & \ \ p_4 \ \ & \
\ p_5 \ \ & \ \ p_6 \ \ & \ \
p_7 \ \ & \ \ p_8 \ \ & \ \ p_9 \ \ \\
\hline
X_{32}^1 & 1 & 0 & 0 & 0 & 0 & 1 & 0 & 1 & 0  \\
X_{32}^2 & 1 & 0 & 0 & 0 & 0 & 0 & 1 & 1 & 0  \\
X_{24}^1 & 0 & 1 & 1 & 0 & 0 & 0 & 0 & 0 & 1  \\
X_{24}^2 & 0 & 1 & 0 & 1 & 0 & 0 & 0 & 0 & 1  \\
X_{31}^1 & 1 & 1 & 1 & 0 & 0 & 0 & 0 & 0 & 0  \\
X_{31}^2 & 1 & 1 & 0 & 1 & 0 & 0 & 0 & 0 & 0  \\
X_{14}^1 & 0 & 0 & 0 & 0 & 0 & 1 & 0 & 1 & 1  \\
X_{14}^2 & 0 & 0 & 0 & 0 & 0 & 0 & 1 & 1 & 1  \\
X_{43}^{11} & 0 & 0 & 1 & 0 & 1 & 1 & 0 & 0 & 0  \\
X_{43}^{12} & 0 & 0 & 0 & 1 & 1 & 1 & 0 & 0 & 0  \\
X_{43}^{21} & 0 & 0 & 1 & 0 & 1 & 0 & 1 & 0 & 0  \\
X_{43}^{22} & 0 & 0 & 0 & 1 & 1 & 0 & 1 & 0 & 0
\end{array}\right)
\label{P_Q222}
\eeq}
From it, we read the matrix implementing the F-term constraints

{\footnotesize
\beq
Q_F=\left(\begin{array}{ccccccccc}
\ \ p_1 \ \ & \ \ p_2 \ \ & \ \ p_3 \ \ & \ \ p_4 \ \ & \
\ p_5 \ \ & \ \ p_6 \ \ & \ \
p_7 \ \ & \ \ p_8 \ \ & \ \ p_9 \ \ \\
\hline
1 & -2 & 1 & 1 & 0 & -1 & -1 & 0 & 1  \\
0 & -1 & 1 & 1 & 0 & -1 & -1 & 1 & 0  \\
0 & 1 & -1 & -1 & 1 & 0 & 0 & 0 & 0
\end{array}\right)
\label{QF_Q222}
\eeq}

Quiver charges are given by

{\footnotesize
\beq
\begin{array}{c|ccccccccc}
\ \ \ \ \ & \ \ p_1 \ \ & \ \ p_2 \ \ & \ \ p_3 \ \ & \ \ p_4 \ \ & \
\ p_5 \ \ & \ \ p_6 \ \ & \ \
p_7 \ \ & \ \ p_8 \ \ & \ \ p_9 \ \ \\
\hline
\ \ Q_1 \ \ &  0 & -1 & 0 & 0 &  0 & 0 & 0 & 0 &  1 \\
\ \ Q_2 \ \ & -1 &  1 & 0 & 0 &  0 & 0 & 0 & 0 &  0  \\
\ \ Q_3 \ \ &  1 &  0 & 0 & 0 & -1 & 0 & 0 & 0 &  0  \\
\ \ Q_4 \ \ &  0 &  0 & 0 & 0 &  1 & 0 & 0 & 0 & -1
\end{array}
\label{Q_Q222}
\eeq}

We consider CS levels $k=(1,1,-1,-1)$. Then, we can impose the $Q_1+Q_3$ and $Q_2+Q_3$ D-terms

{\footnotesize
\beq
Q_D=\left( \begin{array}{ccccccccc}
\ \ p_1 \ \ & \ \ p_2 \ \ & \ \ p_3 \ \ & \ \ p_4 \ \ & \
\ p_5 \ \ & \ \ p_6 \ \ & \ \
p_7 \ \ & \ \ p_8 \ \ & \ \ p_9 \ \ \\
\hline
1 & -1 & 0 & 0 & -1 & 0 & 0 & 0 & 1 \\
0 & 1 & 0 & 0 & -1 & 0 & 0 & 0 & 0
\end{array}
\right)
\label{QD_Q222}
\eeq}

The toric diagram is obtained as the kernel of $Q_{tot}=(Q_F,Q_D)$, and is given
by
{\footnotesize
\beq
G^T=\left( \begin{array}{ccccccccc}
\ \ p_1 \ \ & \ \ p_2 \ \ & \ \ p_3 \ \ & \ \ p_4 \ \ & \
\ p_5 \ \ & \ \ p_6 \ \ & \ \
p_7 \ \ & \ \ p_8 \ \ & \ \ p_9 \ \ \\
\hline
-1 & 0 & 0 & 0 & 0 & 0 & 0 & 0 & 1 \\
2 & 1 & 2 & 0 & 1 & 2 & 0 & 1 & 0 \\
0 & 0 & 0 & 0 & 0 & -1 & 1 & 0 & 0 \\
0 & 0 & -1 & 1 & 0 & 0 & 0 & 0 & 0
\end{array}
\right)
\label{G_Q222}
\eeq}
We can drop the second row and plot the toric diagram. The result is
$\mathcal{C}(Q^{2,2,2})$ toric diagram in \fref{toric_Q222_0}, with multiplicity 3 for the GLSM fields associated with the node at the center.

As for phase I, we see the full $SU(2)^3$ symmetry of $Q^{2,2,2}$. GLSM fields on opposite corners of the toric diagram form doublets according to

{\footnotesize
\beq
\begin{array} {c|ccccc}
& SU(2)_1 & \ \ \ \ & SU(2)_2 & \ \ \ \ & SU(2)_3 \\ \hline
\ \ \ (p_6,p_7) \ \ \ & \fund & & & &  \\
\ \ \ (p_3,p_4) \ \ \ & & & \fund & &  \\
\ \ \ (p_1,p_9) \ \ \ & & & & & \fund
\end{array}
\label{SU2_p}
\eeq}

\bigskip
\bigskip
%%%%%%%%%%%%%%%%%%%%%%%%%%%%%%%%%%%%%%%%%%%%%%%%%%%%%%%%%%%%%%%%%%%%%%%%%%%%%%%%

\end{document}